\def\em{\rm e}

\def\citerefp#1{(\citeauthor{#1}~\citeyear{#1}, and references therein)}
\def\para{\parallel}
\def\citeAY#1{\citeauthor{#1}~\citeyear{#1}}
\def\citeA#1{\citeauthor{#1}}
\def\citeY#1{\citeyear{#1}}

\documentclass{aa}

\usepackage{epsfig,amssymb,amsmath,multirow}
\usepackage{aas_macros,longtable,color}
\usepackage{times,graphicx}
\usepackage{mathrsfs}
\usepackage{float}
\usepackage{txfonts}
\usepackage{natbib}

\usepackage[english,frenchb]{babel}

\usepackage[latin1]{inputenc}		

\begin{document}
\def\abstractname{Abstract}
\renewcommand{\refname}{References}

\newcommand\SgrA{\ensuremath{\mathrm{Sgr~A^\ast}}}
\newcommand\Ek{\ensuremath{E_{k}}}
\newcommand\bfr[1]{\textcolor{red}{\bf #1}}
\newcommand\bfb[1]{\textcolor{blue}{\bf #1}}
\newcommand\bfg[1]{\textcolor{green}{\bf #1}}
\newcommand\itbfg[1]{\textcolor{green}{\it #1}}

\newcommand{\beq}{\begin{equation}}
\newcommand{\bea}{\begin{eqnarray}}
\newcommand{\eeq}{\end{equation}}
\newcommand{\eea}{\end{eqnarray}}
\newcommand{\pe}{\perp}
\newcommand{\pa}{\parallel}
\newcommand{\p}{\partial}


\title{Positron transport in the interstellar medium}

\author{P.~Jean\inst{1}
        \and W.~Gillard\inst{1,2,3} 
	\and A.~Marcowith\inst{4}
	\and K.~Ferri\`ere\inst{5}
}

\offprints{P.~Jean,~\email{Pierre.Jean@cesr.fr}}

\institute{
$^{1}$ CESR, Universit\'e de Toulouse, CNRS, INSU : 9, avenue du colonel Roche, BP 44346, 31028 Toulouse, FRANCE \\
$^{2}$ KTH, Department of Physics, AlbaNova University Centre, SE-106 91Stockholm, SWEDEN \\
$^{3}$ The Oskar Klein Centre for Cosmo Particle Physics, AlbaNova, SE7-106 91Stockholm, SWEDEN \\
$^{4}$ LPTA, CNRS, Universit\'e Montpellier II, Montpellier, FRANCE \\
$^{5}$ LATT, Universit\'e de Toulouse, CNRS: 14, avenue \'Edouard Belin, 31400 Toulouse, FRANCE \\
}

\date{Received  ; accepted }
 
\titlerunning{Positron transport in the ISM}
\authorrunning{P.~Jean, W.~Gillard, A.~Marcowith \& K.~Ferri\`ere.}


\abstract
{}
{We seek to understand the propagation mechanisms of positrons in 
the interstellar medium (ISM). This understanding is a key to 
determine whether the spatial distribution of the annihilation 
emission observed in our Galaxy reflects the spatial distribution 
of positron sources and, therefore, makes it possible to
place constraints on the origin of positrons.}
{We review the different processes that are likely to affect the 
transport of positrons in the ISM. These processes fall into three broad 
categories: scattering off magnetohydrodynamic waves, collisions with 
particles of the interstellar gas and advection with large-scale fluid 
motions. We assess the efficiency of each process and describe its impact 
on the propagation of positrons. We also develop a model of positron 
propagation, based on Monte-Carlo simulations, which enable us to estimate 
the distances traveled by positrons in the different phases of the ISM.}
{We find that low-energy ($\lesssim10\rm~MeV$) positrons 
generally have negligible interactions with magnetohydrodynamic waves, 
insofar as these waves are heavily damped. Positron propagation is 
mainly controlled by collisions with gas particles. 
Under these circumstances, positrons can travel huge distances 
(up to $\sim 30{\rm~kpc}/n_{\rm H,cm^{-3}}$ for 1 MeV positrons) 
along magnetic field lines before annihilating.}
{}


\maketitle


\section{Introduction}

Positron annihilation in the Galactic center (GC) region 
is now a firmly established source of radiation, which has been 
observed since the early 
seventies in several balloon and satellite experiments 
(see \citeAY{2003-von-Ballmoos_397}, ~\citeAY{2004-Jean_552}, 
\citeAY{2006-Diehl_NP777} for reviews). Despite significant 
progress in observational capabilities, the origin of Galactic 
positrons remains an open question.

Recent observations of the 511 keV line intensity using the SPI 
spectrometer onboard the International Gamma-Ray Laboratory 
(INTEGRAL) observatory have revealed a diffuse emission, distributed in 
the bulge and the disk of our Galaxy 
\citep{Knodlseder:2005qy,Weidenspointner:2006lr}. Observations further
indicate that the bulge-to-disk luminosity ratio
(hereafter B/D) of the 511 keV line is rather large ($\approx 3 - 9$)
compared to the distribution of any candidate source. Under the
hypothesis that positrons annihilate close to their sources, the 
spatial distribution of the annihilation emission should reflect 
the spatial distribution of the positron sources. In this view, the large 
B/D ratio could be explained by sources belonging to 
the old stellar population \citep{Knodlseder:2005qy} and a disk 
emission, partly or totally, attributed to the radioactive 
decay of $^{26}$Al and $^{44}$Ti produced in massive stars. The most 
recent SPI analysis covering more than 4 years of data shows hints 
of a longitudinal asymmetry in the spatial distribution of
the 511 keV line emission produced in the 
inner part of the Galactic disk \citep{2008-Weinden-511-Assym1}. 
A similar asymmetry is observed in the distribution of low-mass 
X-ray binaries emitting at high energies, suggesting that these objects
might be the dominant sources of positrons. This conclusion is 
contingent upon the hypothesis that positrons annihilate close to their 
sources, a hypothesis that should be called into question.

Several authors argued that positrons annihilate in the vicinity 
of their production sites, based on the assumption that particles 
propagate according to the so-called Bohm diffusion, i.e., with a 
mean free-path equal to their Larmor radius \citep{Boehm2004, Wang2006}. 
This implicitly supposes that magnetic fluctuations in the ISM 
are strongly tangled on all scales, as is likely the case near strong 
shocks in supernova remnants. In reality, 
given the complexity and the variety of thermodynamical phases 
in the ISM, Bohm diffusion probably 
overestimates particle confinement \citep{Parizot2004}.
\cite{Jean:2006fk} estimated the distances traveled by positrons using 
a model combining a  quasi-linear diffusion theory of wave-particle 
interactions at high energy and an approximate propagation model 
including collisions
with the ambient ISM matter at low energy. A preliminary model of the gas 
content in the Galactic bulge (a more complete description of the gas 
spatial distribution in this region is presented in \citeAY{Ferriere_2007}) 
was then added to the propagation model.  The authors concluded that MeV 
positrons injected by radioactive processes into the bulge cannot escape 
from it and that a single source releasing positrons in the bulge 
might have difficulties accounting for the observed spatial extent of 
the annihilation emission. \cite{2006-Prantzos_449} proposed a solution 
to explain the large B/D ratio measured with SPI, in which 
positrons produced by type Ia supernovae in the old disk are 
transported along magnetic field lines into the bulge. \cite{2006-Cheng_645}
suggested that positrons in the bulge could originate from 
the decay of $\pi^+$ produced in high-energy $pp$ collisions. In this work, 
the energetic protons would be shock-accelerated when a star is tidally disrupted 
in the accretion disk of Sagittarus A$^\ast$ (\SgrA). Since high-energy 
($>30$~MeV) positrons take as long as $\sim 10^7$ years to cool down, they 
should be able to propagate far away from \SgrA{} and to fill the Galactic 
bulge. In order to explain the annihilation emission in the bulge, 
\cite{2006-Totani_P58} proposed a scenario in which MeV positrons were 
produced in the accretion disk of \SgrA{} $\sim 10^7$ years ago. These 
positrons would have filled the bulge while being transported by 
large-scale outflows ($\sim 100$ km s$^{-1}$).  All these studies invoked 
positron transport to explain the observed annihilation emission, but 
they did not include any detailed examination of the physical processes 
that could contribute to it. 

More recently, \cite{Higdon_2009} discussed 
the role of magnetohydrodynamical (MHD) fluctuations in positron 
transport in the different ISM phases. They further estimated the positron 
mean free-path using a phenomenological model of MeV electron transport 
in interplanetary turbulence, especially in the ionized phases where 
turbulence is undamped down to small scales (see appendix).  However, 
the model used to derive the positron mean free-path is strictly valid 
for interplanetary plasmas, and the nature of ISM turbulence 
is likely different. The model developed by \cite{Higdon_2009} is 
up to now the most valuable attempt to account for the propagation of 
low-energy positrons in the ISM. Here, rather than considering a particular 
turbulent model, we review the transport processes that govern 
the positron mean free-path. 

We discuss {\it three} different regimes of positron transport in 
the Galaxy, namely, the regime dominated by scattering off MHD waves 
(section~\ref{sec:MHD}), the regime dominated by collisions with 
particles of the interstellar gas (section~\ref{sec:COL}) and a 
regime of advection with large-scale fluid motions (section~\ref{S:Lsf}).
Based on the quantitative results obtained in each section, we infer 
the relevance and the importance of the respective regimes. 
In section~\ref{sec:App}, we summarize our study and discuss its possible 
implications for positron transport at Galactic scales.


\section{Scattering off magnetohydrodynamic waves}
\label{sec:MHD}

In section \ref{S:Res}, we present the condition for wave-particle
resonance, restricting our analysis to magnetohydrodynamic (MHD) 
waves. We further discuss the effective value of the Alfv\'en speed in the
neutral phases of the ISM. In section \ref{S:Damping}, we present the properties of the MHD wave 
cascades initiated at large spatial scales, e.g., by the 
differential rotation of the Galaxy or by the explosion of supernovae.
In section \ref{S:Cas}, we present the different energy transfer times 
associated with the different type of cascades that likely 
developped in the ISM. In section \ref{S:Sum}, we discuss the collisional 
and collisionless damping mechanisms of MHD waves as well 
as the smallest scales of the MHD cascades in the different ISM phases. 
Only a brief summary is given there; the technical derivation is postponed 
to Appendix \ref{S:Appen}. In section \ref{S:Int}, we discuss the {\it global} 
effect on positron  transport of MHD waves injected at large scales. 
In section \ref{S:Wav}, we consider the {\it local} effect of magnetic 
fluctuations generated by the streaming instability, which can 
survive down to the small scales of low-energy positrons.

Throughout section~\ref{sec:MHD}, we assume, for simplicity, 
that the interstellar gas contains only hydrogen.

\subsection{Positron resonance with MHD waves}
\label{S:Res}

In a medium with non-vanishing magnetic field, wave-particle 
interactions proceed through the Landau-synchrotron resonance 
condition expressed as \citep{Melrose86}:
\beq
\label{Eq:Lsr}
\omega - k_{\pa} \, \upsilon_{\pa} = \ell \, \Omega_{\rm se} \ ,
\eeq
where $\omega$ is the wave frequency, $k_{\pa}$ the component of the 
wave vector along magnetic field lines, $\upsilon_{\pa}$ the particle 
velocity along field lines, $\ell$ an integer, and $\Omega_{\rm se} = 
\Omega_{\rm ce}/\gamma$ the synchrotron frequency of the particle (a 
positron e$^+$ in the case at hand, hence the subscript e), with 
$\gamma$ the particle Lorentz factor, $\Omega_{\rm ce} = (e B / 
m_{\rm e} c)$ the particle cyclotron frequency, and $B$ the magnetic 
field strength. 

In this work, we consider only MHD waves, more specifically, shear
Alfv\'en waves and fast magnetosonic waves as these waves are a major part of the
magnetic fluctuations that pervade the ISM \citep{Lithwickgoldreich01}. We do not consider slow
waves separately, because the dynamics of the slow wave cascade were
shown to be entirely controlled by the Alfv\'en wave cascade
\citep{Lithwickgoldreich01} and the slow wave spectrum is basically 
the same as the Alfv\'en wave spectrum. Following \cite{Yanlazarian04}, we
assume that the transport of charged particles is governed
by either Alfv\'en waves or fast magnetosonic waves. Note that 
higher-frequency waves are potentially important as well, as they can 
easily fulfill the above resonance condition with positrons produced 
by radioactive decay (energy $\lesssim1\rm~MeV$). 
However, whistler waves, the most interesting waves in this frequency
domain, are right-handed polarized and, therefore, cannot be in 
resonance with positrons. \citet{Higdon_2009} have discussed
the possible role of self-generated whistler waves. We will come back on this 
possibility in section \ref{S:Wav}. The possible effects of large-scale compressible 
motions will be examined in section \ref{S:Lsf}.

For the waves of interest here, it can be shown that $\omega \ll 
\Omega_{\rm se}$,\footnote{According to Eq.~(\ref{Eq:thr_omega}), 
this condition is automatically satisfied for positrons with $\gamma 
\ll m_{\rm p} / m_{\rm e}$, corresponding to energies $\ll 
0.94$~GeV.} so we are entitled to neglect $\omega$ in 
Eq.~(\ref{Eq:Lsr}), except in the particular case $\ell = 0$. The 
case $\ell = 0$ corresponds to the so-called Cherenkov resonance, in 
which the particle interacts with a zero-frequency wave in a frame 
moving at velocity $\upsilon_{\pa}$. In that case, Eq.~(\ref{Eq:Lsr}) 
reduces to $\omega = k_{\pa} \upsilon_{\pa}$. The Cherenkov 
resonance can be important for magnetosonic waves, which have a 
perturbed magnetic field component parallel to the background 
magnetic field.

Here, we consider only the dominant harmonics $\ell = \pm 1$. With
$\omega \ll \Omega_{\rm se}$, Eq.~(\ref{Eq:Lsr}) can then be recast 
into the form
\beq
\label{Eq:Lsi}
k_{\pa} \, r_{\rm L}' \, \cos \alpha \simeq \pm 1 \ ,
\eeq
where $\alpha$ is the particle pitch angle (angle between the 
particle velocity and the local magnetic field), $r_{\rm L}' = 
(\upsilon / \Omega_{\rm se}) = (p c / e B)$ is the particle 
gyroradius divided by $\sin \alpha$ (simply referred to as the 
gyroradius in the following), and $p = \gamma m_{\rm e} \upsilon$ is 
the particle momentum. Eq.~(\ref{Eq:Lsi}) expresses the resonance condition in terms of the 
particle gyroradius. It can be rewritten in terms of the particle momentum as
\beq
\label{Eq:Lsp}
p = \pm \frac{eB}{c\,k_\para\,\cos\alpha}
\eeq
or in terms of the particle kinetic energy, $\Ek$, as
\beq
\label{Eq:LsEk}
\Ek= (m_{\rm e} \, c^2) \ \left[\sqrt{\left(\frac{eB}{m_{\rm 
e}c^2\,k_\para\,\cos\alpha}\right)^2+1}-1\right] \ .
\eeq
\centerline{* * * * *}
\bigskip

Roughly speaking, MHD waves can exist only at frequencies lower than 
the proton cyclotron frequency, $\Omega_{\rm cp} = (e B / m_{\rm p} 
c)$:
\beq
\label{Eq:thr_omega}
\omega \le \Omega_{\rm cp} \ .
\eeq
Either they are damped by collisional effects (mainly viscous 
friction and ion-neutral collisions) at low frequencies (see 
section~\ref{S:Col}) or, if they manage to survive collisional 
effects, then at frequencies approaching $\Omega_{\rm cp}$, they are 
heavily damped by Landau damping due to thermal protons (see 
section~\ref{S:Nco}).

For Alfv\'en waves ($\omega = V_{\rm A} \, k_{\pa}$, with $V_{\rm A}$ 
the Alfv\'en speed), Eq.~(\ref{Eq:thr_omega}) is equivalent to
\beq
\label{Eq:thr_kpar_alfven}
k_{\pa} \le \frac{\Omega_{\rm cp}}{V_{\rm A}} \ ,
\eeq
which, in view of the resonance conditions, Eqs.~(\ref{Eq:Lsp}) and
(\ref{Eq:LsEk}), implies a threshold on the positron momentum 
(obtained for $|\cos \alpha| = 1$)~:
\beq
\label{Eq:thr_mom_alfven}
p \ge m_{\rm p} \, V_{\rm A} \ ,
\eeq
and a threshold on the positron kinetic energy~:
\beq
\label{Eq:thr_KE_alfven}
\Ek \ge (m_{\rm e} \, c^2) \
\left[
  \sqrt{
    \left( \frac{m_{\rm p} \, V_{\rm A}}
                {m_{\rm e} \, c}
    \right)^2 + 1
       } - 1
\right] \ .
\eeq

For fast magnetosonic waves ($\omega = V_{\rm F} (\Theta) \, k$, with 
$\Theta$ the angle between the wave vector and the magnetic field, 
$V_{\rm F}$ the phase speed of the fast mode, comprised between 
$V_{\rm F}(0) = {\rm max} (V_{\rm A},C_{\rm s})$ and $V_{\rm F}(\pi / 
2) = V_{\rm ms} \equiv \sqrt{V_{\rm A}^2 + C_{\rm s}^2}$, and $C_{\rm 
s}$ the adiabatic sound speed), Eq.~(\ref{Eq:thr_omega}) leads to $k 
\le \Omega_{\rm cp} / V_{\rm F} (\Theta)$ and, hence, $p \ge m_{\rm 
p} \, V_{\rm F} (\Theta) / \cos \Theta$. Since the functions $V_{\rm 
F} (\Theta)$ and $V_{\rm F} (\Theta) / \cos \Theta$ reach their 
minimum values at $\Theta = 0$, the absolute requirement on the 
wavenumber reads
\beq
\label{Eq:thr_k_fast}
k \le \frac{\Omega_{\rm cp}}{V_{\rm F} (0)} \ ,
\eeq
and the thresholds on the positron momentum and kinetic energy (again 
obtained
for $|\cos \alpha| = 1$) are given by
\beq
\label{Eq:thr_mom_fast}
p \ge m_{\rm p} \, V_{\rm F}(0)
\eeq
and
\beq
\label{Eq:thr_KE_fast}
\Ek \ge (m_{\rm e} \, c^2) \
\left[
  \sqrt{
    \left( \frac{m_{\rm p} \, V_{\rm F}(0)}
                {m_{\rm e} \, c}
    \right)^2 + 1
       } - 1
\right] \ ,
\eeq
respectively.\\

\begin{table*}[!t]
\centering
\begin{tabular}{lccccccc}
\hline
\hline
ISM phase  &  
$T~{\rm [K]}$  &  $B~{\rm [\mu G]}$  & $n_{\rm H}~{\rm [cm^{-3}]}$  &  $f_{\rm ion}$  & $n_{\rm i}~{\rm [cm^{-3}]}$  & $k_{\pa{\rm max}}~{\rm [10^{-9}~cm^{-1}]}$  & $E_{\rm k,min}~{\rm [keV]}$
\\
\hline\\
HIM (low $B$)  & $10^6$  &  2  &  $0.005 - 0.01$  &  1  &  $0.005 - 0.01$  & $3.1 - 4.4$  & $35 - 18$  \\
HIM (high $B$)  & $10^6$  &  20  &  $0.005 - 0.01$  &  1  &  $0.005 - 0.01$  & $3.1 - 4.4$  & $1500 - 950$  \\
WIM  & 8000  &  5  &  $0.2 - 0.5$  &  $0.6 - 0.9$  &  $0.12 - 0.45$  & $15 - 29$  & $9.5 - 2.5$  \\
WNM  &  $6000 - 10000$  &  5  &  $0.2 - 0.5$   &  $0.007 - 0.05$  & $0.0014 - 0.025$  & $1.6 - 6.9$  &  $540 - 45$  \\
CNM  &  $50 - 100$  &  6  &  $20 - 50$  &  $4 \times 10^{-4} - 10^{-3}$  & $0.008 - 0.05$  & $3.9 - 9.8$  &  $175 - 32$  \\
MM  &  $10 - 20$  &  $8.5 -850$  &  $10^2 - 10^6$  & $\lesssim 10^{-4}$  &  & $\lesssim 4.4$  &  $\gtrsim 265$  \\
\hline
\end{tabular}
\caption{Physical parameters of the different ISM phases: molecular 
medium (MM), cold neutral medium (CNM), warm neutral medium (WNM), warm 
ionized medium (WIM) and hot ionized medium (HIM). $T$ is the 
temperature, $B$ the magnetic field strength, $n_{\rm H}$ the 
hydrogen density, $f_{\rm ion} = n_{\rm i} / (n_{\rm i} + n_{\rm n})$ 
the ionization fraction, $n_{\rm i}$ the ion density, $k_{\pa{\rm 
max}}$ the maximum parallel wavenumber of Alfv\'en waves (right-hand 
side of Eq.~(\ref{Eq:thr_kpar_bis})), and $E_{\rm k,min}$ the minimum 
kinetic energy required for positrons to interact resonantly with 
Alfv\'en waves (right-hand side of Eq.~(\ref{Eq:thr_KE_bis})). Here, 
we assume a pure-hydrogen gas, for which $n_{\rm i} = f_{\rm ion} \, 
n_{\rm H}$ and $n_{\rm n} = (1 - f_{\rm ion}) \, n_{\rm H}$.
}
\label{T:Res}
\end{table*}

The question now is what expression should be used for the Alfv\'en 
speed. In a fully ionized medium, the Alfv\'en speed is simply 
$V_{\rm A} = B / \sqrt{4 \pi \rho_{\rm i}}$, with $\rho_{\rm i}$ the 
ion mass density. However, in a partially ionized medium, the 
relevant Alfv\'en speed depends on the degree of coupling between 
ions and neutrals. If the ion-neutral and neutral-ion collision 
frequencies, $\nu_{\rm in}$ and $\nu_{\rm ni}$, are much greater than 
the wave frequency, $\omega$, then ions and neutrals are very well 
coupled together through ion-neutral collisions, and as a result, an 
Alfv\'en wave will set the entire fluid (ions + neutrals) into 
motion. In that case, one should use the total Alfv\'en speed, 
$V_{\rm A,tot} = B / \sqrt{4 \pi \rho_{\rm tot}}$, with $\rho_{\rm 
tot} = \rho_{\rm i} + \rho_{\rm n}$ the total (ion + neutral) mass 
density. In contrast, if the ion-neutral and neutral-ion collision 
frequencies are much smaller than the wave frequency, then ions and 
neutrals are no longer coupled together, and an Alfv\'en wave will 
solely set the ions into motion. In that case, one should use the 
ionic Alfv\'en speed, $V_{\rm A,i} = B / \sqrt{4 \pi \rho_{\rm i}}$, 
as in a fully ionized medium.

In an atomic medium with temperature $T \lesssim 100$~K, the 
ion-neutral and neutral-ion collision frequencies are given by 
$\nu_{\rm in} \simeq (1.6 \times 10^{-9}~{\rm cm^3~s^{-1}}) \, n_{\rm 
n}$ and 
$\nu_{\rm ni} \simeq (1.6 \times 10^{-9}~{\rm cm^3~s^{-1}}) \, n_{\rm 
i}$, where $n_{\rm n}$ and $n_{\rm i}$ are the neutral and ion number 
densities, respectively \citep{1961ApJ...134..270O}. At high 
temperature, the collision frequencies increase as $\sqrt{T}$ 
\citep{1965-Braginskii_RPP1,1987-Shull_134}; assuming an effective 
cross section for H-H$^+$ collisions $\sim 10^{-14}~{\rm cm^2}$ 
\citep{Wentzel74}, we find that, for $T \gtrsim 140$~K, $\nu_{\rm in} 
\sim (1.4 \times 10^{-9}~{\rm cm^3~s^{-1}}) \, \sqrt{T/100~{\rm K}} 
\, n_{\rm n}$ and $\nu_{\rm ni} \sim (1.4 \times 10^{-9}~{\rm 
cm^3~s^{-1}}) \, \sqrt{T/100~{\rm K}} \, n_{\rm i}$. In a molecular 
medium, the collision frequencies are $\nu_{\rm in} \simeq (2.1 
\times 10^{-9}~{\rm cm^3~s^{-1}}) \, n_{\rm n}$ and $\nu_{\rm ni} 
\simeq (2.1 \times 10^{-9}~{\rm cm^3~s^{-1}}) \, n_{\rm i}$ 
\citep{1961ApJ...134..270O}. For comparison, the frequency of a 
resonant Alfv\'en wave is given by $\omega = V_{\rm A} \, k_{\pa}$ 
with, according to Eq.~(\ref{Eq:Lsi}), $k_{\pa} \gtrsim 1 / r_{\rm 
L}' = (e B / p c)$, i.e.,
$\omega \gtrsim (6.6 \times 10^{-5}~{\rm s}^{-1}) \, \left( B_{\rm 
\mu G}^2 / \sqrt{n_{\rm cm^{-3}}} \right) \, (p c)_{\rm MeV}^{-1}$, 
where $n$ is the relevant number density (e.g., $n = n_{\rm i} + 
n_{\rm n}$ if $\nu_{\rm in},\nu_{\rm ni} \gg \omega$ and $n = n_{\rm 
i}$ if $\nu_{\rm in},\nu_{\rm ni} \ll \omega$). From this, it follows 
that for typical interstellar conditions (see Table~\ref{T:Res}), 
$\nu_{\rm in},\nu_{\rm ni} \ll \omega$ up to positron energies of at 
least 1~GeV. Hence, in the present context, the relevant Alfv\'en 
speed is $V_{\rm A,i} = B / \sqrt{4 \pi \rho_{\rm i}}$, not only in 
the ionized phases, but also in the so-called neutral (i.e., atomic 
and molecular) phases.
\bigskip

\noindent \underline{Alfv\'en waves}
\medskip

With the above statements in mind, the requirement on the parallel 
wavenumber of Alfv\'en waves, Eq.~(\ref{Eq:thr_kpar_alfven}), can be 
rewritten as
\beq
\label{Eq:thr_kpar_bis}
k_{\pa} \, \le \,
\frac{\sqrt{4 \pi} \, e}{\sqrt{m_{\rm p}} \, c} \ \sqrt{n_{\rm i}}
\, = \, 
(4.4 \times 10^{-8}~{\rm cm}^{-1}) \ \sqrt{n_{\rm i,cm^{-3}}} \ .
\eeq
The corresponding condition on the positron kinetic energy, 
Eq.~(\ref{Eq:thr_KE_alfven}), becomes
\beq
\label{Eq:thr_KE_bis}
\Ek \ge (511~{\rm keV}) \ 
\left[ 
  \sqrt{
    1.8 \times 10^{-4} \
    \frac{B_{\rm \mu G}^2}{n_{\rm i,cm^{-3}}}
    + 1
       } - 1
\right] \ .
\eeq
The above expressions were obtained on the assumption that the only 
ion present in the ISM is H$^+$. To account for the presence of other 
ions, it suffices, in good approximation, to replace $n_{\rm i}$ by 
$n_{\rm H} + 4 \, n_{\rm He}$ with $n_{\rm He} \simeq 0.1 \, n_{\rm 
H}$ in the hot phase and by $n_{\rm H^+} + 12 \, n_{\rm C,gas}$ with 
$n_{\rm C,gas} \simeq 1.4 \times 10^{-4} \, n_{\rm H}$ 
\citep{1996-Cardelli_467} in the atomic phases.

The maximum parallel wavenumber, $k_{\pa{\rm max}}$, and the minimum 
kinetic energy, $E_{\rm k,min}$, given by the right-hand sides of 
Eqs.~(\ref{Eq:thr_kpar_bis}) and (\ref{Eq:thr_KE_bis}), respectively, 
are listed in Table~\ref{T:Res} for the different phases of the ISM. 
Also listed in Table~\ref{T:Res} are the estimated temperature, $T$, 
magnetic field strength, $B$, hydrogen density, $n_{\rm H}$, and 
ionization fraction, $f_{\rm ion} = n_{\rm i} / (n_{\rm i} + n_{\rm 
n})$ of the different phases. The values of $T$, $n_{\rm H}$ and 
$f_{\rm ion}$ are taken from the review paper of 
\cite{2001-Ferriere_RMP73}.
For $B$, we adopt the value of $5~{\rm \mu G}$ inferred from rotation 
measure studies \citep[e.g.][]{1989-Rand_343,1993-Ohno_M262} for the 
warm phases, the value of $6~{\rm \mu G}$ inferred from Zeeman 
splitting measurements \citep{2005-Heiles_624} for the cold phase, 
the relation $B \propto \sqrt{n_{\rm H}}$ normalized to $B = 85~{\rm 
\mu G}$ at $n_{\rm H} = 10^4~{\rm cm^{-3}}$ 
\citep{1999ApJ...520..706C} for molecular clouds, and the two extreme 
values of $2~{\rm \mu G}$ and $20~{\rm \mu G}$ for the hot phase. The 
lower value pertains to the standard scenario in which the hot gas is 
generated by stellar winds and supernova explosions, which sweep up 
the ambient magnetic field lines and evacuate them from the hot 
cavities. The higher value pertains to an alternative scenario in 
which large-scale highly turbulent MHD fluctuations produce magnetic 
fields above equipartition with the local thermal pressure 
(\citeAY{Bykov01}; \citeAY{Parizotetal04}; and references therein).\\

From Table~\ref{T:Res}, it emerges that the maximum parallel 
wavenumber of Alfv\'en waves is typically a few $10^{-9}~{\rm 
cm^{-1}}$, close to the largest wavenumber, $k_{\rm L} \simeq 
10^{-9}~{\rm cm^{-1}}$, of the electron density power spectrum 
inferred from interstellar scintillation \citep{Armstrongetal95}. 
Furthermore, the minimum kinetic energy required for positrons to 
interact resonantly with Alfv\'en waves varies from a few keV (in the 
warm ionized medium) to a few hundreds of keV (in regions with large 
Alfv\'en speeds, namely, in molecular clouds and possibly in the warm 
neutral and hot ionized media). For comparison, positrons produced by 
radioactive decay are injected into the ISM with typical kinetic 
energies $\sim 1$~MeV. This means that positrons from radioactive 
decay can interact resonantly with Alfv\'en waves only over a 
restricted energy range. This range is particularly narrow in regions 
with large Alfv\'en speeds, such as molecular clouds (where $V_{\rm  
A,i} \gtrsim 185~{\rm km s}^{-1}$); it can even vanish in the hot phase if 
the magnetic field is as strong as $20~{\rm \mu G}$ (implying $V_{\rm 
A,i} \gtrsim 436~{\rm km s}^{-1}$). In contrast, the resonant range 
extends over at least two orders of magnitude in the warm ionized 
medium, where the ion density is highest (and $V_{\rm A,i} \simeq (16 
- 31)~{\rm km s}^{-1}$). The reason why resonant interactions with 
Alfv\'en waves are no longer possible below $E_{\rm k,min}$ is 
because the Larmor radius has become smaller than the smallest 
possible scale of existing Alfv\'en waves.
\bigskip

\noindent \underline{Fast magnetosonic waves}
\medskip

The numerical expressions and values of the wavenumber and kinetic 
energy thresholds can be obtained in the same manner as for Alfv\'en 
waves, with the two following differences: First, the wavenumber 
threshold (Eq.~(\ref{Eq:thr_k_fast})) applies to the total 
wavenumber, as opposed to the parallel wavenumber. Second, the speed 
entering the expressions of the thresholds is the phase speed of the 
fast mode for parallel propagation, $V_{\rm F}(0)$, as opposed to the 
Alfv\'en speed, $V_{\rm A}$. In practice, however, the second 
difference is only formal, except in the hot low-$B$ phase. Indeed, 
$V_{\rm F}(0) = {\rm max} (V_{\rm A},C_{\rm s})$, and ${\rm max} 
(V_{\rm A},C_{\rm s}) = V_{\rm A}$ in all the ISM phases, except in 
the hot low-$B$ phase, where ${\rm max} (V_{\rm A},C_{\rm s}) = 
C_{\rm s} \simeq 166~{\rm km s}^{-1}$. In consequence, the values of the 
maximum wavenumber, $k_{\rm max}$, and the minimum kinetic energy, 
$E_{\rm k,min}$, are those listed in Table~\ref{T:Res}, except in the 
hot low-$B$ phase, for which Eqs.~(\ref{Eq:thr_k_fast}) and 
(\ref{Eq:thr_KE_fast}) lead to $k_{\rm max} \simeq 1.2 \times 
10^{-9}~{\rm cm^{-1}}$ and $E_{\rm k,min} \simeq 220~{\rm keV}$, 
respectively. The latter value is roughly an order of magnitude 
larger than for Alfv\'en waves, which significantly narrows down the 
energy range over which positrons from radioactive decay can interact 
resonantly with fast magnetosonic waves.

The Cherenkov resonance (for $\ell =0$) occurs when $\omega = k_{\pa}
\, \upsilon_{\pa}$, i.e., at a wave propagation angle $\Theta_{\rm 
T}$ such that $V_{\rm F}(\Theta_{\rm T}) = \upsilon_{\pa} \, \cos 
\Theta_{\rm T}$, independent of the wavenumber. Since $V_{\rm 
F}(\Theta_{\rm T})$ never departs from the fast magnetosonic speed, 
$V_{\rm ms}$, by more than a factor $\sqrt{2}$, this expression is 
approximately equivalent to $\cos \Theta_{\rm T} = V_{\rm ms} / 
\upsilon_{\pa}$. The Cherenkov resonance requires that the wave at 
propagation angle $\Theta_{\rm T}$ exist and not be damped by a 
collisional or collisionless process (see next subsections).
\bigskip

Let us re-emphasize that the above results should be considered only 
as rough estimates. As we will now see, both Alfv\'en and fast 
magnetosonic waves are subject to various damping processes in the 
ISM.

\subsection{MHD wave cascades}
\label{S:Damping}

\subsubsection{Energy transfer timescales}
\label{S:Cas}
We consider both Alfv\'en waves and fast magnetosonic waves as parts 
of turbulent cascades. The main sources of turbulence able to 
counterbalance the dissipation mechanisms expected in the ISM are 
the magnetorotational instability driven by the differential rotation 
of the Galaxy and the explosion of supernovae \citep{2004RvMP...76..125M}. 
Both mechanisms release energy at {\it large scales}. In the rest of the 
paper, we adopt $L_{\rm inj} = 100$~pc for the injection scale. 
This does not preclude the possibility of injecting magnetic 
fluctuations at smaller scales; an issue discussed in section \ref{S:Wav}.
\bigskip

\noindent \underline{Alfv\'en wave cascade}
\medskip

The most recent developments in MHD turbulence theory explain the 
energy cascade towards smaller scales by the distortion of oppositely 
travelling Alfv\'en wave packets 
\citep[e.g.,][]{Lithwickgoldreich01}. The kinematics of the 
interactions produce a highly anisotropic cascade, which 
redistributes most of the energy in the perpendicular 
scales.\footnote{Throughout this paper, the perpendicular and 
parallel directions are taken with respect to the mean magnetic field 
direction.} 
We will return to this important question in section \ref{S:Int}.

The transfer time of the Alfv\'en wave cascade, $\tau_{\rm A}$, 
corresponds to the wave packet crossing time along the mean magnetic 
field:
\beq
\label{Eq:transfer_alfven}
\tau_{\rm A} = \frac{1}{V_{\rm A} \, k_{\pa}} \ .
\eeq
For reference, when $L_{\rm inj} = 100$~pc, one has
$$
\label{Eq:transfer_alfven_num}
\tau_{\rm A} = (1.4 \times 10^{15}~{\rm s}) \ 
\frac{\sqrt{n_{\rm cm^{-3}}}}{B_{\rm \mu G}} \
\frac{1}{k \, L_{\rm inj} \, \cos \Theta} \ ,
$$
where $n$ is the relevant number density (in a fully ionized medium, 
$n = n_{\rm i}$, while in a partially ionized medium, $n = n_{\rm i} 
+ n_{\rm n}$ if $\nu_{\rm in},\nu_{\rm ni} \gg \omega$ and $n = 
n_{\rm i}$ if $\nu_{\rm in},\nu_{\rm ni} \ll \omega$).
\bigskip

\noindent \underline{Fast magnetosonic wave cascade}
\medskip

The transfer time of the fast magnetosonic wave cascade, $\tau_{\rm 
F}$, can be written as
\beq
\label{Eq:transfer_fast}
\tau_{\rm F} 
\, = \, 
\frac{\omega}{k^2 \, \delta \upsilon_k^2}
\, = \, 
\frac{V_{\rm F} \, L_{\rm inj}^{1/2}}
     {\delta \upsilon_{\rm inj}^2} \ k^{-1/2} \ ,
\eeq
where $\delta \upsilon_k$ is the turbulent velocity at scale 
$k^{-1}$, $\delta \upsilon_{\rm inj}$ is the turbulent velocity at 
the injection scale, $L_{\rm inj}$, and, as before, $V_{\rm F}$ is 
the phase speed of the fast mode \citep{Yanlazarian04}. In writing 
the second identity, we assumed a Kraichnan spectrum ($dE/dk \propto 
k^{-3/2}$ or $\delta \upsilon_k \propto k^{-1/4}$). For our numerical 
estimates, we adopt again $L_{\rm inj} = 100$~pc and we set both 
$\delta \upsilon_{\rm inj}$ and $V_{\rm F}$ to the fast magnetosonic 
speed, $V_{\rm ms} = \sqrt{V_{\rm A}^2 + C_{\rm s}^2}$ (as mentioned 
earlier, $V_{\rm F}$ does not depart from $V_{\rm ms}$ by more than a 
factor $\sqrt{2}$), whereupon we find for $L_{\rm inj} = 100$~pc
$$
\label{Eq:transfer_fast_num}
\tau_{\rm F} = (1.4 \times 10^{15}~{\rm s}) \
\frac{\sqrt{n_{\rm cm^{-3}}}}
     {B_{\rm \mu G} \sqrt{1 + C_{\rm s}^2 / V_{\rm A}^2}} \
(k \, L_{\rm inj})^{-1/2} \ ,
$$
with the density $n$ defined as for the Alfv\'en wave cascade.
\bigskip

For both Alfv\'en and fast magnetosonic waves, the dominant damping 
process depends on the wavelength (or inverse wavenumber) compared to 
the proton collisional mean free-path, 
\beq
\label{Eq:Lmfp}
\lambda_{\rm p} = \upsilon_{\rm p} \ \tau_{\rm p} 
\simeq (3.5 \times 10^{13}~{\rm cm}) \ 
\frac{T_{\rm p,eV}^2}{n_{\rm e,cm^{-3}} \, \Lambda} \ ,
\eeq
where $\upsilon_{\rm p} = \sqrt{3 \, k_{\rm B} T_{\rm p} / m_{\rm 
p}}$ is the proton r.m.s. velocity, $\tau_{\rm p}$ the proton 
collision time, $n_{\rm e}$ the electron density, and $\Lambda$ the 
Coulomb logarithm, given by 
$$
\Lambda = 23.4 - 0.5 \, \ln n_{\rm e,cm^{-3}} + 1.5 \, \ln T_{\rm 
e,eV} \ , 
\quad {\rm for} \ T_{\rm e} < 50~{\rm eV}
$$
and 
$$
\Lambda = 25.3 - 0.5 \, \ln n_{\rm e,cm^{-3}} + \ln T_{\rm e,eV} \ , 
\quad \quad \ \ {\rm for} \ T_{\rm e} > 50~{\rm eV}
$$
\citep{1965-Braginskii_RPP1}. Waves with $k^{-1} > \lambda_{\rm p}$ 
are basically collisional and, therefore, affected by collisional 
damping (see section~\ref{S:Col}), whereas waves with $k^{-1} < 
\lambda_{\rm p}$ are basically collisionless and affected by 
collisionless damping (see section~\ref{S:Nco}).

As we will see below, in all the cases considered here, $L_{\rm inj} 
> \lambda_{\rm p}$, which means that both turbulent cascades start in 
the collisional range. The wave energy is then transferred to smaller 
scales up to the point where the wave damping rate, $\Gamma$, becomes 
equal to the transfer rate. In other words, the turbulent cascades 
are cut off at a wavenumber
$k_{\rm cut}$ such that
\beq
\label{Eq:cutoff_alfven}
\Gamma \ \tau_{\rm A} = 1
\eeq
for the Alfv\'en cascade and
\beq
\label{Eq:cutoff_fast}
\Gamma \ \tau_{\rm F} = 1
\eeq
for the fast magnetosonic cascade. If the collisional damping rate, 
$\Gamma_{\rm coll}$, is high enough that $\Gamma_{\rm coll} \, 
\tau_{\rm A} = 1$ at a scale larger than $\lambda_{\rm p}$, then the 
Alfv\'en cascade is cut off by collisional damping at that scale. On 
the other hand, if $\Gamma_{\rm coll} \, \tau_{\rm A} < 1$ down to 
$\lambda_{\rm p}$, then the Alfv\'en cascade proceeds down to the 
collisionless range and is eventually cut off by collisionless 
damping at a scale smaller than $\lambda_{\rm p}$. Similarly for the 
fast magnetosonic cascade.

\subsubsection{Damping and cutoff}
\label{S:Sum}

In this section, we summarize the investigation presented in Appendix 
\ref{S:Appen} on the dominant damping processes of the Alfv\'en and 
fast magnetosonic wave cascades in the different ISM phases.

In the mostly neutral, atomic and molecular phases of the ISM, 
the Alfv\'en and fast magnetosonic wave cascades are, regardless of 
their origin, both cut off by ion-neutral collisions at scales larger 
than the proton mean free-path, $\lambda_{\rm p}$, i.e.,
in the collisional regime -- and thus at scales considerably 
larger 
than the Larmor radii of interstellar positrons.
In consequence, positrons will find no Alfv\'en or fast magnetosonic 
waves from an MHD cascade to resonantly interact with (see also \citet{Higdon_2009}).

The situation is completely different in the ionized phases of the 
ISM. There, the Alfv\'en wave cascade develops with insignificant 
(collisional) damping down to $\lambda_{\rm p}$. It then enters the 
collisionless range, where it 
is eventually cut off by linear Landau damping around the proton 
inertial length. Thus, the extended inertial range of the Alfv\'en 
wave 
cascade leaves some room for possible resonant interactions with 
positrons. The fast magnetosonic wave cascade, for its part, suffers 
strong collisional (viscous) damping. In the warm ionized phase, this 
damping is sufficient to kill the cascade (with the possible 
exception of quasi-parallel waves) before it enters the collisionless 
range. In the hot ionized phase, the cascade manages to reach the 
collisionless range, but it is then quickly destroyed by linear 
Landau damping (again with the possible exception of quasi-parallel 
waves). 
Altogether, no fast magnetosonic waves from an MHD cascade have 
sufficiently small scales to come into resonant interactions with 
positrons.

\subsection{Positron interactions with MHD wave cascades}
\label{S:Int}
Our previous discussion indicates that the MHD cascades are truncated
at scales several orders of magnitude larger than the Larmor radii 
of positrons produced by radioactive decay or as cosmic rays, except 
in the hot and warm ionized phases of the ISM. In these phases, Alfv\'en 
wave turbulence is expected to cascade nearly undamped down to scales 
close to the Larmor radius of MeV positrons. However, as we now argue, this 
does not necessarily mean that short-wavelength Alfv\'en waves will resonantly 
interact with MeV positrons. Indeed, magnetic fluctuations are probably highly 
anisotropic at small scales, in the sense that turbulent eddies are strongly 
elongated along the mean magnetic field, or, in mathematical terms, 
$k_{\perp} \gg k_{\parallel}$ \citep{Goldreichsridhar95, Yanlazarian04}.
Because of the important anisotropy of magnetic fluctuations, which 
increases towards smaller scales, scattering off Alfv\'en waves appears 
questionable. The elongated irregularities associated with anisotropic 
turbulence average out over a particle gyration \citep{Chandran00}.
If some Alfv\'en waves are present at scales $\sim r_{\rm L}'$, 
the scattering frequency is reduced by more than 20 orders of  magnitude 
compared to the situation with isotropic turbulence
(see, for instance, \citeAY{Yanlazarian02}). Slab turbulence gives the 
same order of estimates \citep{Yanlazarian04}. In consequence, scattering 
off Alfv\'en wave turbulence should be extremely inefficient at confining 
positrons in the ionized phases of the ISM. In this view, the diffusion models 
for positron transport adopted in a series of recent papers by 
\citet{Parizotetal05,2006-Cheng_645,Jean:2006fk}, using diffusion coefficients 
derived from quasi-linear theory, overestimate the confinement by plasma waves.

To circumvent this problem, \citet{Yanlazarian04} reconsidered scattering 
off fast magnetosonic waves, emphasizing the isotropy of the 
fast wave cascade. As we saw earlier, in the hot and warm ionized phases,
fast waves decay away at scales much larger than the Larmor radii 
of positrons from radioactivity, except possibly at quasi-parallel 
propagation. If propagation angles are not or only weakly randomized 
by wave-wave interactions or by chaotic divergence of magnetic field lines, 
quasi-parallel waves may survive down to much smaller scales.
In that case, they may be involved either in gyroresonance or in 
Cherenkov resonance, also known as transit-time damping (TTD) resonance, with 
positrons.  As explained at the end of section \ref{S:Res}, Cherenkov resonance 
occurs at propagation angles $\Theta_{\rm T}$ such that
$\cos \Theta_{\rm T} \simeq V_{\rm ms} / (\upsilon \, \cos \alpha)$,
independent of the wavelength. But we know that only those fast waves with $\Theta_{\rm T} \simeq 0$
have a chance to escape heavy damping. From this, we conclude that only positrons with pitch angles 
satisfying $\cos \alpha \simeq V_{\rm ms} / \upsilon \ll 1$ have a chance to
experience TTD resonance. At precisely $\cos \alpha = V_{\rm ms} / \upsilon$, the TTD mechanism 
vanishes, but its rate rises rapidly as $\cos \alpha$ increases 
above $V_{\rm ms}/\upsilon$ \citep{1998-Schlickeiser_492}.
Ultimately, positrons produced by radioactive decay are unlikely to 
efficiently interact with MHD waves from a direct cascade 
generated at large scales. Such positron-wave interactions appear to be 
completely ruled out in the neutral phases of the ISM. In the ionized phases, they could 
potentially take place, but only under very restrictive conditions, including quasi-parallel
fast waves ($\Theta_{\rm T} \simeq 0$) and nearly perpendicular positron
motion ($\alpha \simeq \arccos (V_{\rm ms} / \upsilon)$).  \\
Of course, an additional local injection of MHD waves at much smaller scales 
could participate in the confinement of positrons. Some aspects of this possibility will be discussed in section 
\ref{S:Reac}.

\subsection{Wave injection through plasma instabilities}
\label{S:Wav}

As we saw in the previous subsection, MHD waves injected at large scales 
into an MHD cascade are generally unable to efficiently interact with
positrons from radioactive decay. However, MHD waves can be injected 
into the ISM by a variety of fluid or kinetic instabilities, which 
involve changes over the whole or a fraction of the velocity distribution 
of some particle population. These waves can be injected at scales 
$\ll L_{\rm inj}$, possibly directly into the collisionless regime. 
Each case requires a dedicated investigation of the wave damping process. 
The scale of the turbulence injection is a key parameter controlling 
the interaction between MHD waves and positrons. First, $L_{\rm inj}$ 
controls the anisotropy of the Alfv\'en wave cascade at the scale of 
the Larmor radius of positrons. Second, it also controls the cutoff 
wavenumber and the propagation angle of the fast magnetosonic cascade 
(see Eq.~(\ref{Eq:kcut_landau_fast})). In this subsection, we 
focus on some particular aspects of one type of kinetic instability.

One of the most widely studied kinetic instabilities is triggered by 
the streaming of cosmic rays in the ISM, with a bulk velocity larger 
than a few times the local Alfv\'en speed \citep{Wentzel74, Skilling75}. 
The streaming instability is expected to develop mainly in the 
intercloud medium. Cosmic-ray streaming compensates for the sink in the 
low-energy cosmic-ray population due to strong ionization losses inside 
molecular clouds. Low-energy cosmic rays scatter off their self-generated waves, 
and are, therefore, excluded from molecular clouds \citep{Lerche67,
Skillingstrong76, Cesarskyvoelk78, Dogielsharov85}. This scenario was adapted to the 
transport of cosmic-ray electrons by \citet{Morfill82}. The streaming 
instability and other kinds of kinetic instabilities 
recently received new attention in the context of cosmic-ray diffusion 
in anisotropic MHD turbulence \citep{Farmergoldreich04, Lazarianberesnyak06}.

Here, we restrict our discussion to the sole streaming instability. 
The waves are generated at scales $k_{\rm st}^{-1}$ close to the 
gyroradii of low-energy cosmic rays, so that $k_{\rm st}^{-1} \ll L_{\rm inj}$. 
If the waves generated by cosmic rays are to serve as scattering 
agents for low-energy positrons, then the Landau-synchrotron resonance 
condition has to be fulfilled by both species, i.e., by virtue of 
Eq.~(\ref{Eq:Lsp}):
\beq
\label{Eq:Bpe}
p_{\rm e} = p_{\rm p} \ \frac{|\cos\alpha_{\rm p}|}{|\cos\alpha_{\rm 
e}|} \ .
\eeq
Eq.(\ref{Eq:Bpe}) implies that the wave-generating cosmic rays and
the scattered positrons must have comparable momenta, unless the 
ratio of angular factors is very different from unity.
Now, the wave-generating cosmic rays have typical momenta in the range
$[p_{\rm min}, p_{\rm max}]$. Their maximum momentum is set by the 
condition that waves can indeed be amplified by cosmic-ray streaming, which 
leads to $p_{\rm max} \sim 100~{\rm MeV} / c$ \citep{Skillingstrong76}.
Their minimum momentum is set by the highest possible frequency of 
MHD waves, exactly as for positrons, and is therefore given by 
$p_{\rm min} \simeq m_{\rm p} \, V_{\rm A}$ (see 
Eq.~(\ref{Eq:thr_mom_alfven})). This means that, in the warm ionized phase, positrons with kinetic 
energies below $\simeq 2.5~-~10$ keV (depending on the exact ion density) can 
no longer resonate with Alfv\'en waves generated by cosmic-ray streaming 
(see Table \ref{T:Res}). We conclude that this mechanism does not operate over a broad range
of positron momenta and that it probably does little to confine positrons 
within the ionized phases of the ISM.

\cite{Higdon_2009} proposed an alternative mechanism whereby positrons 
scatter off their own self-generated waves. The growth rate of such an 
instability is proportional to the density of resonant particles. 
The positron density depends strongly on their position with respect 
to the sources. \citet{Higdon_2009} argued that whistler waves are
heavily damped in the neutral phases of the ISM, but the streaming 
instability may also operate close to the sources (for instance 
close to supernova remnants; see Ptuskin et al. 2008) or above 
the Galactic disk where the ionization fraction can be close to unity. 
A complete estimation of this process deserves a detailed investigation
and will be considered in a forthcoming paper.

\subsection{Positron transport in dissipated turbulence and re-acceleration}
\label{S:Reac}

In the solar wind, MeV positrons can also interact with perturbations 
that fall into the dissipative range of the turbulence, above the 
steepening observed at a fraction of the proton gyrofrequency at 
wavenumbers $k \gtrsim k_{\rm A}$. It has been proposed that low-frequency 
{\it non-resonant magnetosonic} waves can dominate the propagation of 
sub-MeV particles if magnetosonic waves are present in the solar wind 
\citep{Toptyghin1985,Ragot2006}.
\cite{1999-Ragot_518} suggested that non-resonant fast magnetosonic 
waves can produce efficient angular scattering through pitch angles 
$\alpha~\rightarrow~\pi/2$ and thus govern the particle mean 
free-path at energies $\sim 1$~MeV. This result could be applied
to the stellospheres of massive stars (about a few parsecs) 
and to larger regions of the ISM (see discussion in \citet{Higdon_2009},
though in a different perspective).

The electric component of the waves induces a variation 
in particle energy. The average effect of such interactions 
leads to a stochastic energy gain, also known as second-order 
Fermi acceleration.
Particle stochastic acceleration (also referred to as re-acceleration) 
is important 
if the re-acceleration time is shorter than the energy loss time,
i.e., if $t_{\rm stoch} \lesssim t_{\rm loss}$.
The re-acceleration time is related to the momentum diffusion coefficient 
$D_p$ through $t_{\rm stoch} = p^2/D_p$ (here, we consider
only pitch-angle averaged quantities). 
It can also be expressed in terms of the particle angular diffusion
frequency $\nu_{\rm s}$ as
$t_{\rm stoch} \simeq (c/V_{\rm A})^2 \, \nu_{\rm s}^{-1}$.
Then the above timescale ordering is equivalent to
\beq
\label{Eq:nusf2}
\nu_{\rm s} \gtrsim \left(\frac{c}{V_{\rm A}}\right)^2 
\ \frac{1}{t_{\rm loss}} \ .
\eeq
Particles with energies $\lesssim 1$ GeV mostly suffer ionization 
and Coulomb losses. 
The energy loss rate depends on whether the medium is ionized or neutral. 
\citet{Ginzburg-1979bs} gives
\beq
\frac{1}{t_{\rm loss}}
= \frac{3}{4} \ \sigma_{\rm T} \, c \, n_{\rm e} \ 
\frac{73.6 + \ln(\gamma/n_{\rm e, cm^{-3}})}{\beta (\gamma -1)}
\eeq
in an ionized medium and
\beq
\frac{1}{t_{\rm loss}}
= \frac{3}{4} \, \sigma_{\rm T} \, c \, n_{\rm H^0} \ 
\frac{20.5 + \ln \left[ (\gamma - 1) (\gamma^2 - 1) \right]}{\beta 
(\gamma -1)}
\eeq
in a neutral medium, where $\sigma_{\rm T}$ is the Thomson 
cross-section, $\beta = \upsilon/c$
and, as before, $\gamma$ is the particle Lorentz factor,
$n_{\rm e}$ the free-electron density
and $n_{\rm H^0}$ the density of hydrogen atoms.
The condition (\ref{Eq:nusf2}) translates into an upper limit 
on the mean free-path:
\beq
\label{Eq:lamio}
\lambda \lesssim (3.4 \times 10^{-5} {\rm pc}) \
\left(\frac{B_{\rm \mu G}^2}{n_{\rm e, cm^{-3}} \, n_{\rm i, cm^{-3}}} \right) 
\ \frac{(\gamma - 1) \beta^2}{\ln(\gamma/n_{\rm e, cm^{-3}})+73.6}
\eeq
in an ionized medium and
\beq
\label{Eq:lamne}
\lambda \lesssim (3.4 \times 10^{-5} {\rm pc}) \
\left(\frac{B_{\rm \mu G}^2}{n_{\rm H^0, cm^{-3}} \, n_{\rm i, cm^{-3}}}\right) 
\ \frac{(\gamma - 1) \beta^2}{\ln\left[ (\gamma - 1) (\gamma^2 - 1) 
\right]+20.5}
\eeq
in a neutral medium.
This implies that if, in a particular medium, the angular 
diffusion frequency is high enough, particles are reaccelerated. 
In that case, a solution can be obtained with the help of
a diffusion-convection equation. This type of investigation 
is beyond the scope of the current paper and is postponed to a future work. 
However, the above discussion already calls some of the results derived 
by \citet{Higdon_2009} into question, as these authors ignored 
re-acceleration in their analysis. 
What emerges from the above discussion is that re-acceleration 
can be important, especially in low-density ionized phases.
More specifically, if we introduce the parameter values proposed 
by \citet{Higdon_2009} for the Galactic interstellar bulge 
(see their Table 1), we find that positron re-acceleration is important 
in the very hot phase of the inner bulge, and probably in the hot phase 
of the middle and outer bulge. 


\section{Effects of collisions with gas particles}
\label{sec:COL}

In this section, we examine the case when the trajectories of 
positrons in the ISM is only driven by their collisions 
with gas particles while they propagate along a steady state 
magnetic field line, without considering interactions with 
MHD waves, an aspect usually overlooked by the previous analysis. 
During these collisions, high-energy positrons not only lose energy, but 
they also undergo pitch angle scattering. The scattering of 
kinematic parameters of positrons reduces the maximum distance they 
can travel along a straight line in the ISM. 
In section~\ref{sc:MC-Simu}, we review and describe the interaction 
processes between positrons and interstellar matter. 
In section~\ref{sc:Model}, we describe the methods used to calculate 
positron propagation in this so-called 
``collisional regime'', and in section~\ref{sc:MC-ISM}, we present 
the detailed results of our computations for 1 MeV positrons. 
The main results obtained for positrons with initial kinetic energies 
ranging from 1~keV to 10~MeV in typical ISM phases and the case of 
propagation in a turbulent magnetic field are described 
in section \ref{sc:Disc}.

\subsection{Positron interactions with interstellar matter\label{sc:MC-Simu}}

In the absence of collisions, positrons move along magnetic field lines 
in helical trajectories. When positrons interact with gas particles, 
they can either gain or lose energy in elastic and inelastic 
scattering or even annihilate with free or bound electrons.
The energy and pitch angle variations resulting from the 
interaction depend on the energy of the incident positron 
and on the velocity and nature of the target particle.

The positrons that we are studying spend most of their lifetime 
travelling at high energy, with a kinetic energy greater than the 
thermal energy of target particles in the ISM. Consequently, 
we do not account for the propagation of positrons at thermal energy,
but postpone the discussion of this case to section~\ref{sc:Disc}. 

Interactions of positrons with the ionized component of the 
interstellar gas and, at high energy (E $\gtrsim$ 100 MeV), 
with the magnetic field (synchrotron radiation) and 
the interstellar radiation field (inverse Compton scattering) 
are generally considered as continuous processes.  
They are referred to as {\it continuous energy-loss processes} 
and are presented in section~\ref{sc:CNRJ-loss}. In a mostly 
neutral gas, positrons lose a larger fraction of their energy 
when they excite or ionize atoms or molecules. 
Such interactions, which result in a quantified variation 
of the positron energy, are described in section~\ref{sc:bin-col}. 

\subsubsection{Continuous energy-loss processes\label{sc:CNRJ-loss}}

\begin{figure}[!t]
	\includegraphics*[scale=1]{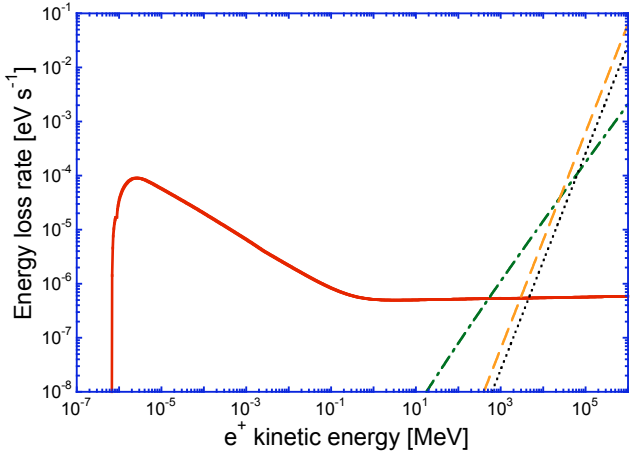}
	\caption{Energy-loss rates of positrons in a 8000~K plasma with 
number densities $n_{\rm e}=n_{\rm H^+}\simeq1\rm~cm^{-3}$. The {\it 
long-dashed line} shows the energy-loss rate due to synchrotron 
radiation assuming $B_0=5\rm~\mu G$ and an average value of 
sin$^2\alpha=2/3$. The {\it dotted line} is the 
energy-loss rate by inverse Compton scattering assuming $U_{\rm 
ph}=0.26\rm~eV~cm^{-3}$, which corresponds to the cosmic microwave 
background energy density. The {\it dot-dashed line} shows the 
energy-loss rate by bremsstrahlung emission due to collisions with 
charged particles. The {\it solid 
line} corresponds to the positron energy-loss rate due to 
Coulomb interactions between positrons and electrons.  
\label{fig:dedt}}
\end{figure}

Figure~\ref{fig:dedt} shows the energy loss rates of positrons as 
functions of their energy in a fully ionized plasma 
with temperature $T= 8000\rm~K$. The energy loss rates by 
synchrotron radiation and inverse Compton scattering are 
derived from \cite{Blumenthal-1970it}. They are proportional to 
the magnetic-field and photon energy densities, respectively. 
The energy loss rate by bremsstrahlung is calculated using the 
approximation presented in \cite{Ginzburg-1979bs}. The energy 
loss rate due to Coulomb collisions (free-free) is calculated using 
the Bhabha cross-section as described in \cite{2007-Asano_A168}.

In a fully ionized plasma, Coulomb collisions represent the 
dominant loss process for positrons with $E \lesssim$ 100 MeV. 
Since the energy loss rates by synchrotron radiation and inverse 
Compton scattering depend on the magnetic-field and photon energy 
densities, they vary with the location of positrons in our Galaxy. 
It is expected that inverse Compton losses become 
dominant in zones close to stellar clusters. 
However, we are interested in positrons 
with kinetic energy $\lesssim$ 10 MeV and, therefore, we
may neglect synchrotron losses as long as $B \lesssim$ 1 mG.

The average scattering angle induced by Coulomb collisions 
occurring over a time interval $\delta t$ is estimated through the 
relation

\begin{equation}
	\sin\bar{\theta}=\sqrt{\frac{d\langle\sin^2\theta\rangle}{dt}\delta t}\quad,
	\label{eq:ScatAngl}
\end{equation}
where $d\langle\sin^2\theta\rangle/dt$ is the rate of variation 
of the average $\sin^2\theta$ and $\theta$ is the positron 
scattering angle in the laboratory frame.
The rates of variation of the average scattering angle are evaluated 
by integrating the Bhabha (e$^+$e$^-$ collision) and Rutherford (e$^+$p 
collision) differential cross sections (see Appendix in 
\citeAY{2007-Asano_A168}), over the scattering angles as

\begin{equation}
	\frac{d\langle\sin^2\theta\rangle}{dt}=\upsilon n 
	\int_{\theta^\star_{\rm min}}^{\pi}\sin^2\theta \ 
\frac{d\sigma^\star}{d\cos\theta^\star}\ \sin\theta^\star d\theta^\star\quad,
	\label{eq:avgSinTheta}
\end{equation}
where $n$ is the number density of target particles, 
$\theta^\star$ is the scattering angle in the center-of-mass frame, 
$\frac{d\sigma^\star}{d\cos\theta^\star}$ 
is the differential cross section of the interaction and  
$\theta^\star_{\rm min}$ is the minimum scattering angle (see Eq.~(B7) in 
\citeAY{Dermer:1985lr}).

While the energy loss rate from e$^+$p collisions is negligible 
compared to that from e$^+$e$^-$ collisions, their deviation rates are 
equivalent. We checked that the resulting deviation rates are in agreement 
with those derived at low energy using the formalism of \cite{NRL-Huba}.

\subsubsection{Annihilation and interactions with atoms and molecules\label{sc:bin-col}}

In a neutral medium, high-energy positrons lose energy mainly 
by ionizing and exciting atoms and molecules, or they annihilate directly 
with bound electrons, the elastic scattering process being negligible 
at these energies (see \citeAY{Charlton:2000lr} and \citeAY{1994-Wallyn_422}). 
Such interactions occur at random while positrons propagate in the ISM. 
Therefore, we will determine the kind of interaction and calculate the 
variation of the kinematic parameters with a Monte-Carlo method 
that incorporates the corresponding cross sections.

The ionization and excitation cross sections as well as the 
differential cross sections as functions of the energy lost
by positrons in ionizing collisions were calculated by 
~\cite{1965-Gryzinski_PR138a,1965-Gryzinski_PR138b,1965-Gryzinski_PR138c}. 
The cross section of annihilation in flight of positrons with bound electrons 
is equal to that of annihilation with free electrons, since the binding energy 
of electrons is negligible with respect to the kinetic energy of the 
positrons under study. We use the cross section of annihilation with 
free electrons presented in \citeA{Guessoum_al_2005} 
(\citeY{Guessoum_al_2005}, and references therein). 

In the Monte-Carlo simulations, the energy lost by a positron when it ionizes 
an atom/molecule is chosen randomly according to its differential cross 
section. The energy lost by a positron when it excites an 
atom/molecule is derived from the energies of the atomic levels involved in 
the interaction. Once the energy lost by ionization or excitation is known, 
the scattering angle of the positron is calculated with the kinematics 
of the interaction, assuming that the atoms/molecules stay at rest 
and assuming azimuthal symmetry.

\subsection{Simulations of the collisional transport \label{sc:Model}}

Our Monte-Carlo simulations are based on the methods presented 
in \cite{1979-Bussard_228} and \cite{Guessoum_al_2005}. However, we 
add the calculation of the trajectories of positrons while they 
propagate along magnetic field lines and include additional steps 
to account for the scattering of positrons by collisions. 

\begin{figure}[!b]
	\centering
	\includegraphics*[scale=0.93]{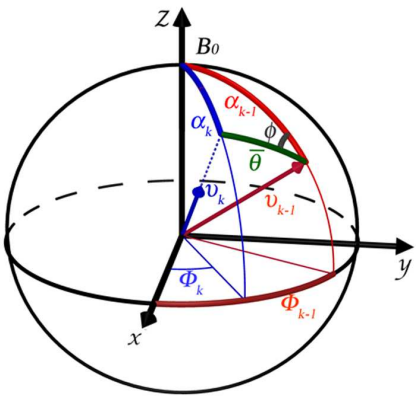}
	\caption{Kinematic parameters of the scattering 
	in the laboratory frame. The magnetic field $\vec{B}_0$ lies along 
	the $z$ axis. $\alpha_{k-1}$ is the pitch angle between the positron 
	velocity $\vec{\upsilon}_{k-1}$ and $\vec{B}_0$ before the 
	interaction. $\alpha_{k}$ is the pitch angle after the interaction, 
	when the positron velocity is $\vec{\upsilon}_{k}$. $\Phi_{k-1}$ 
	and $\Phi_{k}$ are the phases of incident positrons before and 
	after the interaction, respectively. $\bar{\theta}$ is the scattering 
	angle of positrons and $\phi$ is the azimuthal scattering angle. 
\label{fig:repere}}
\end{figure}

At the initial time ($k=0$), positrons are located 
at $(x,y,z,t)=(0,0,0,0)$ in a 
frame such that the magnetic field $\vec{B_0}$ is directed along 
the $z$ axis. The initial kinetic energy of positrons is $E_0$.
The direction of their initial velocity $\vec{\upsilon}_{0}$, 
which is defined by their initial pitch angle $\alpha_{0}$ and their 
initial phase $\Phi_{0}$ (see Figure~\ref{fig:repere}), is chosen randomly 
according to an isotropic velocity distribution in the entire space. 

We proceed through successive iterations. At step $k$, the resulting 
pitch angle ($\alpha_{k}$) and phase ($\Phi_{k}$) are given by  

\begin{equation}
	\label{eq:ep-angle}
	\displaystyle \cos\alpha_{k} \ = \ \displaystyle  \cos\alpha_{k-1} \ 
\cos\bar{\theta}+\sin\alpha_{k-1} \ \sin\bar{\theta} \ \cos\phi \quad,
\end{equation}

\begin{equation}
\cos(\Phi_{k}-\Phi_{k-1}) = \frac{\cos\bar{\theta}-\cos\alpha_{k-1} \ 
\cos\alpha_{k}}{\sin\alpha_{k-1} \ \sin\alpha_{k}}\quad,
\end{equation}
where the azimuthal scattering angle $\phi$ is a random number 
uniformly distributed between 0 and 2$\pi$. The kinetic energy $E_{k}$ 
after collision with an atom/molecule or annihilation with an electron 
is given by  \cite{1979-Bussard_228} :

\begin{equation}
	{\rm R}=\exp\left[-\int_{E_{k-1}}^{E_{k}}\frac{\upsilon \ \sum\limits_{i,j}n_j \ 
\sigma_{i,j}(E)}{dE/dt}~dE\right]\quad,
\end{equation}
where R is a random number uniformly distributed between 0 and 
1, $\sigma_{i,j}$ is the cross section of process $i$ (see  
section \ref{sc:bin-col}) with target $j$ and $n_j$ is the 
number density of target $j$. The kind of interaction is chosen 
randomly according to its probability 

\begin{equation}
	\displaystyle P_{i,j}(\Ek)=\frac{n_{j} \ 
\sigma_{i,j}(\Ek)}{\sum\limits_{i,j}n_{j} \ \sigma_{i,j}(\Ek)}\quad,
\end{equation}
The energy lost and the scattering angle induced by collisions 
are derived using the methods presented in section~\ref{sc:bin-col}.

The increment in time between two iterations, $\delta t$, is equal to 
a scattering time $\delta t_{s}$ defined as

\begin{equation}
    \delta t_{s}= \frac{\epsilon}{d\langle\sin^2\theta\rangle/dt}\quad,
\end{equation}
with $d\langle\sin^2\theta\rangle/dt$ the rate of variation 
of the average $\sin^2\theta$ (see Eq.~\ref{eq:avgSinTheta}) and
$\epsilon$ a prescribed number $\ll 1$. 
During $\delta t_{s}$, positrons lose an amount of energy 
$dE/dt \times \delta t_{s}$ via continuous energy-loss processes, 
and, they are slightly scattered by an angle $\bar{\theta}$ obtained from 
Eq.~\ref{eq:ScatAngl}.
If the time interval between two successive collisions with neutrals
is shorter than the scattering time $\delta t_{s}$ (e.g., in a
weakly ionized medium), then the increment in time $\delta t$ is replaced 
by the time interval between two collisions.

Between two steps and/or two interactions, positrons propagate 
in a regular helical trajectory along magnetic field lines 
with a gyroradius $r_{\rm L}' \, \sin\alpha_{k-1}$ (defined 
in section \ref{S:Res}). The positron position after each step 
is therefore calculated analytically.

The above procedure is repeated until positrons annihilate or their kinetic 
energy falls below 100 eV, since the distance 
traveled below this energy becomes negligible.
To estimate the spatial distribution 
of positrons and their lifetime, we read out 
the positron location and slowing-down time at the end of the track. 
To save CPU time while obtaining sufficiently accurate results,
Monte-Carlo simulations are performed with a number of positrons 
ranging from 5000 to 20000.

To validate our code, we performed several tests for positrons 
with $E_{0} <$ 10 MeV released in different media and 
we compared the results of our simulations with previous work. 
The fractions of positrons annihilating in flight with 
free or bound electrons are in agreement with the results presented 
by \cite{2006-Beacom_PRL97} and \cite{Sizun06}.
At low energy ($E_{0} <$ 1 keV), the fractions of positronium formed in 
flight by charge exchange with atoms and molecules in different media 
are identical to those obtained by \cite{Guessoum_al_2005}.
Once positrons are thermalized, their propagation behaves as
classical diffusion. The distance they travel then is negligible 
compared to the distance traveled in the slowing-down regime, except 
in the hot medium (see section \ref{sc:Disc}). 

\subsection{Results of the simulations for 1 MeV positrons \label{sc:MC-ISM}}

This section presents, as an illustrative example, the 
detailed results of simulations obtained for positrons with 
initial kinetic energy $E_{0}$=$1\rm~MeV$. For this first set of 
simulations, we chose the warm medium, whose temperature is 
intermediate between those of the cold and hot media. The warm medium 
is also the most appropriate medium to study the impact of the 
ionization fraction, which covers the whole possible range between 0 
and 1. We assumed the magnetic field to be uniform, with a strength
$B_0=5\rm~\mu G$.

Since the energy loss rates and the frequency of inelastic interactions 
are proportional to the target density, for given 
abundances and a given ionization fraction, the distance traveled by positrons 
scales as the inverse of the total density. 
Therefore, we performed simulations for $n_\ion{H}{}=1\rm~cm^{-3}$, 
with $n_\ion{H}{}=n_{\rm H^0}+n_{\rm H^+}+2n_{\rm H_2}$ the total 
number density of hydrogen nuclei. Preliminary tests 
of our simulations indicate that the presence of 
ionized or neutral helium with an abundance ratio $\rm He/H\approx 0.1$ 
does not strongly affect positron propagation (differences less 
than 10\%). We also neglected molecular hydrogen, except in the MM
where $\rm H_2$ molecules are the dominant species.

\begin{figure}[!b]
	\centering
	\includegraphics*[scale=0.33]{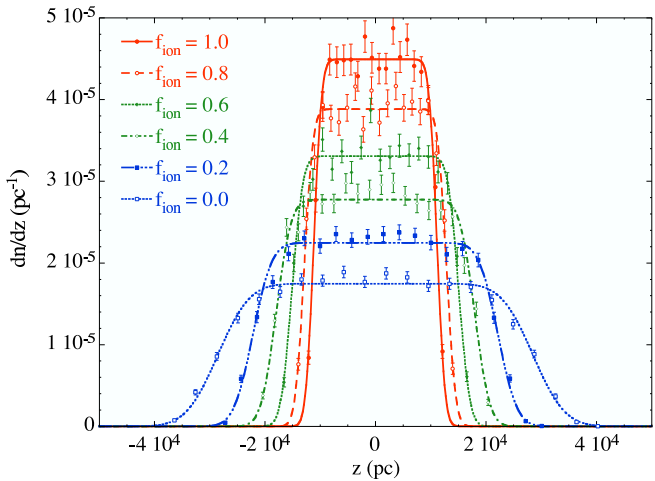}
	\caption{Spatial distributions of positrons along field lines 
	at the end of their slowing-down period in a warm medium ($T$ = 8000 K 
	and $n_\ion{H}{}=1\rm~cm^{-3}$), for several ionization fractions, 
	$f_{\rm ion}$. The distributions are normalized to unity and the 
	error bars are estimated according to Poisson (counting) statistics.   
	\label{fig:dndz-vs-fion}}
\end{figure}

Figure~\ref{fig:dndz-vs-fion} shows the spatial distributions of positrons 
along the magnetic field direction ($z$), as extracted from our simulations, 
once they reach $E =$ 100 eV. Positrons are initially ``injected'' at 
$(x,y,z)=(0,0,0)$ with a pitch angle chosen randomly according 
to an isotropic velocity distribution in the entire space. 
The temperature is 8000 K and the 
ionization fraction ranges from 0 to 1. 
The field-aligned distributions are nearly uniform out to the 
maximum distance traveled along field lines, $d_{\rm max}$ (obtained 
when the pitch angle is and remains equal to 0). This means that the 
pitch angle does not change significantly over most of the 
slowing-down period of positrons. If the pitch angle remained
strictly constant, the distance traveled along field lines 
would be equal to $d_{\rm max} \times \cos\alpha_{0}$. 
Since positrons are initially emitted isotropically, $\cos\alpha_{0}$ 
is uniformly distributed between $-1$ and 1, so that the 
field-aligned distributions of positrons at the end of their 
slowing-down period would be uniform. In reality, the slight pitch 
angle scattering induced by collisions produces a slight scattering of their 
final field-aligned positions and, therefore, smoothes out
the edge of the otherwise uniform distributions.
The maximum distance $d_{\rm max}$ is obtained by integrating over 
energy the ratio of the positron velocity to the energy loss rate. 
In the present case, the energy loss rate is the sum of the 
contributions from Coulomb collisions (see section~\ref{sc:CNRJ-loss}) 
and inelastic interactions (ionization and excitation) with atoms 
and molecules. The latter contributions are evaluated using the 
Bethe-Bloch formula \citep{Ginzburg-1979bs}.

The extent of the spatial distributions in the directions perpendicular 
to the magnetic field is a few times the Larmor radius, 
i.e., negligible with respect to the extent of the field-aligned 
distributions. We characterize the extent of the field-aligned 
distributions presented in Figure~\ref{fig:dndz-vs-fion} by their 
full width at half maximum (FWHM$_{//}$). Its uncertainty is calculated 
with a bootstrap method taking into account the uncertainty in 
the number of positrons (Poisson statistics) in each spatial bin. 

The extent of the field-aligned distributions increases naturally with 
decreasing ionization fraction (see Figure~\ref{fig:dndz-vs-fion}), since 
the energy losses through Coulomb collisions dominate over the losses 
due to inelastic interactions (ionization and excitation) with 
neutrals. Depending on the ionization fraction, the half FWHM$_{//}$ 
is 15\% to 30\% lower than $d_{\rm max}$.

\subsection{Transport of positrons in the different ISM phases\label{sc:Disc}}  

This section summarizes the results of the simulations for
positrons with initial kinetic energies ranging from 1 keV 
to 10~MeV, in the different ISM phases. 

\begin{figure}[!t]
    \centering
    \includegraphics*[scale=0.33]{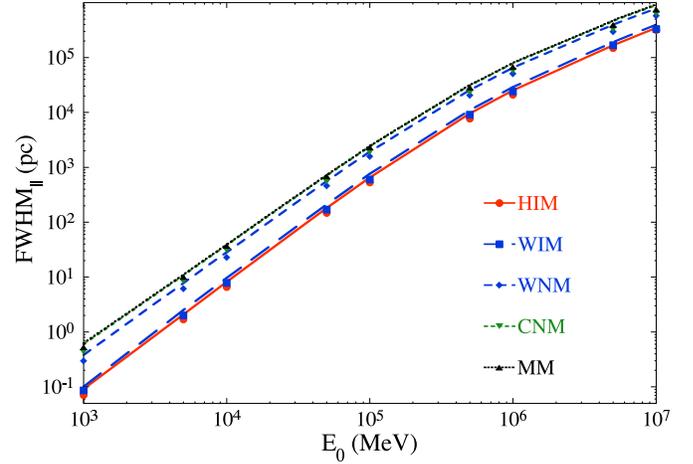}
    \caption{FWHM of the field-aligned distributions of 
at the end of their slowing-down period, as functions of 
their initial kinetic energy. The hydrogen density is arbitrarily 
set to $n_\ion{H}{}=1\rm~cm^{-3}$ in each ISM phase. 
The lines show the distance 2$d_{\rm max}$ (see text for details) while 
points show FWHM$_{//}$ of the distribution as obtained by the simulations.
    \label{fig:FWHM}}
\end{figure}

Figure~\ref{fig:FWHM} shows the extent (FWHM$_{//}$) of the field-aligned 
distribution of positrons at the end of their slowing-down time (i.e., 
when they reach 100 eV), as a function of their initial kinetic energy 
in the different ISM phases. As in section \ref{sc:MC-ISM} , the hydrogen 
density is arbitrarily set to $n_\ion{H}{}=1\rm~cm^{-3}$, and the ionization 
fraction is 0, 0.001, 0.1, 0.9 and 1 in the MM, CM, WNM, WIM and HIM, 
respectively. 
The extents presented in Figure~\ref{fig:FWHM} are compared to the extents 
that the distributions would have if the positron pitch angles remained 
constant, namely, 2$d_{\rm max}$. 
In each ISM phase and for any given initial energy, FWHM$_{//}$ is 
similar to, and always slightly less than, 2$d_{\rm max}$, as expected 
from the effect of pitch angle scattering.

The above FWHM$_{//}$ extents are given assuming a uniform magnetic 
field in the $z$ direction. However, field lines in the 
ISM are perturbed by turbulent motions. As a result, 
realistic magnetic fields in the ISM consist of a mean 
(``regular'') component plus a turbulent component, leading 
to chaotic field lines. The distances traveled by positrons along the 
uniform field (i.e., along the $z$ axis) are shorter than the total 
distances traveled along the actual chaotic field lines.

\begin{figure}[!b]
	\centering	
	\includegraphics*[scale=0.33]{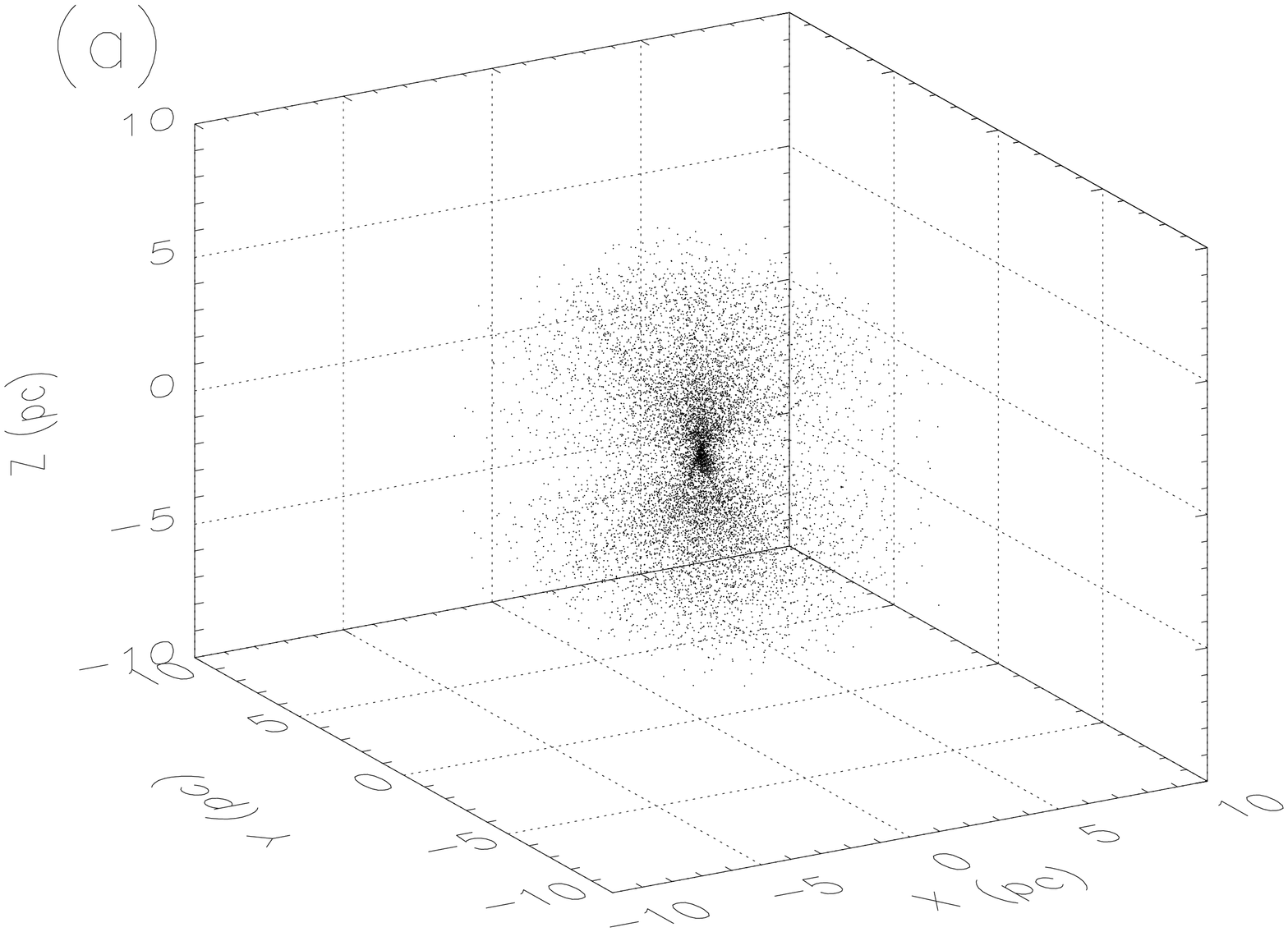}
	\includegraphics*[scale=0.33]{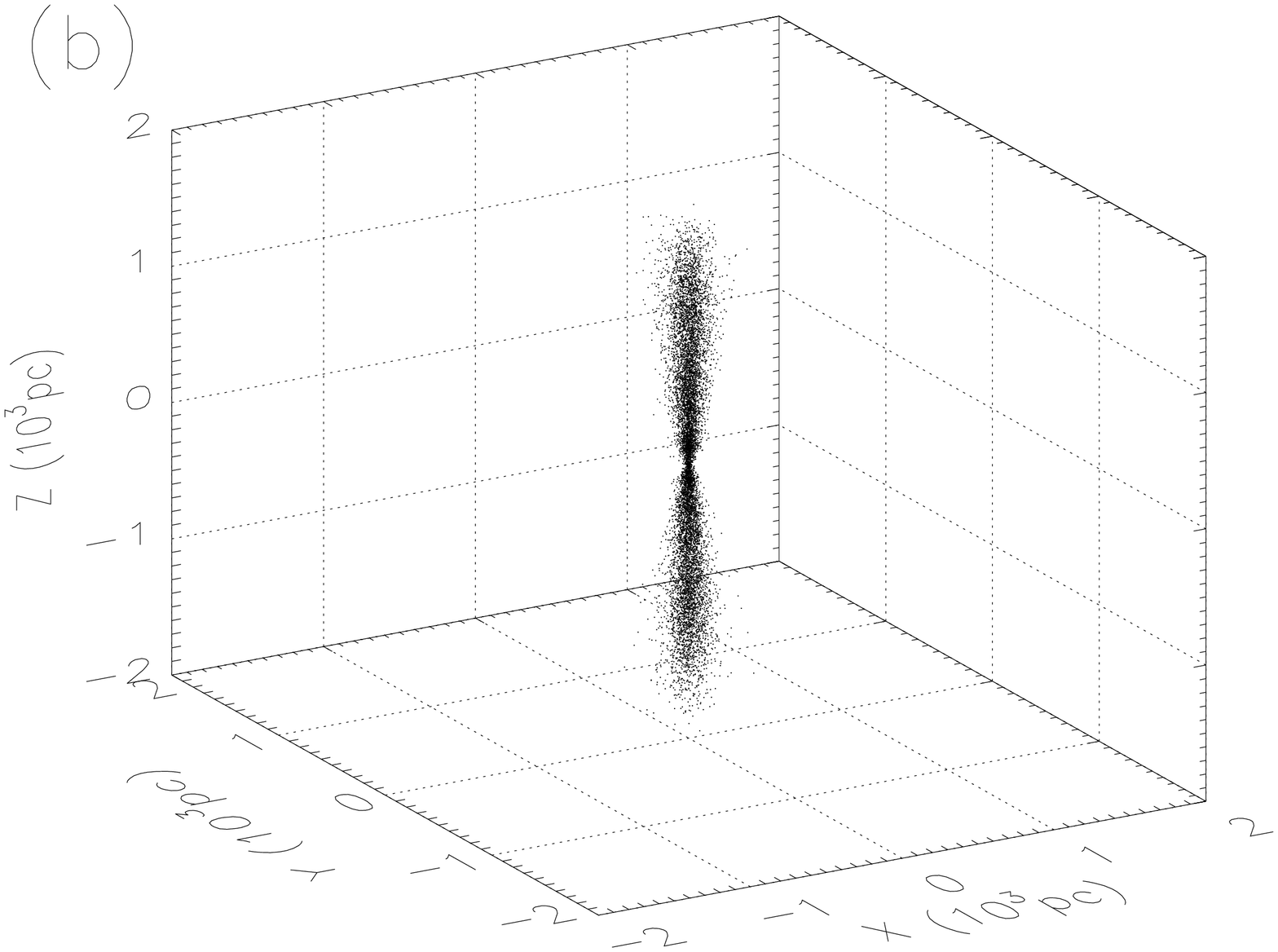}
	\includegraphics*[scale=0.33]{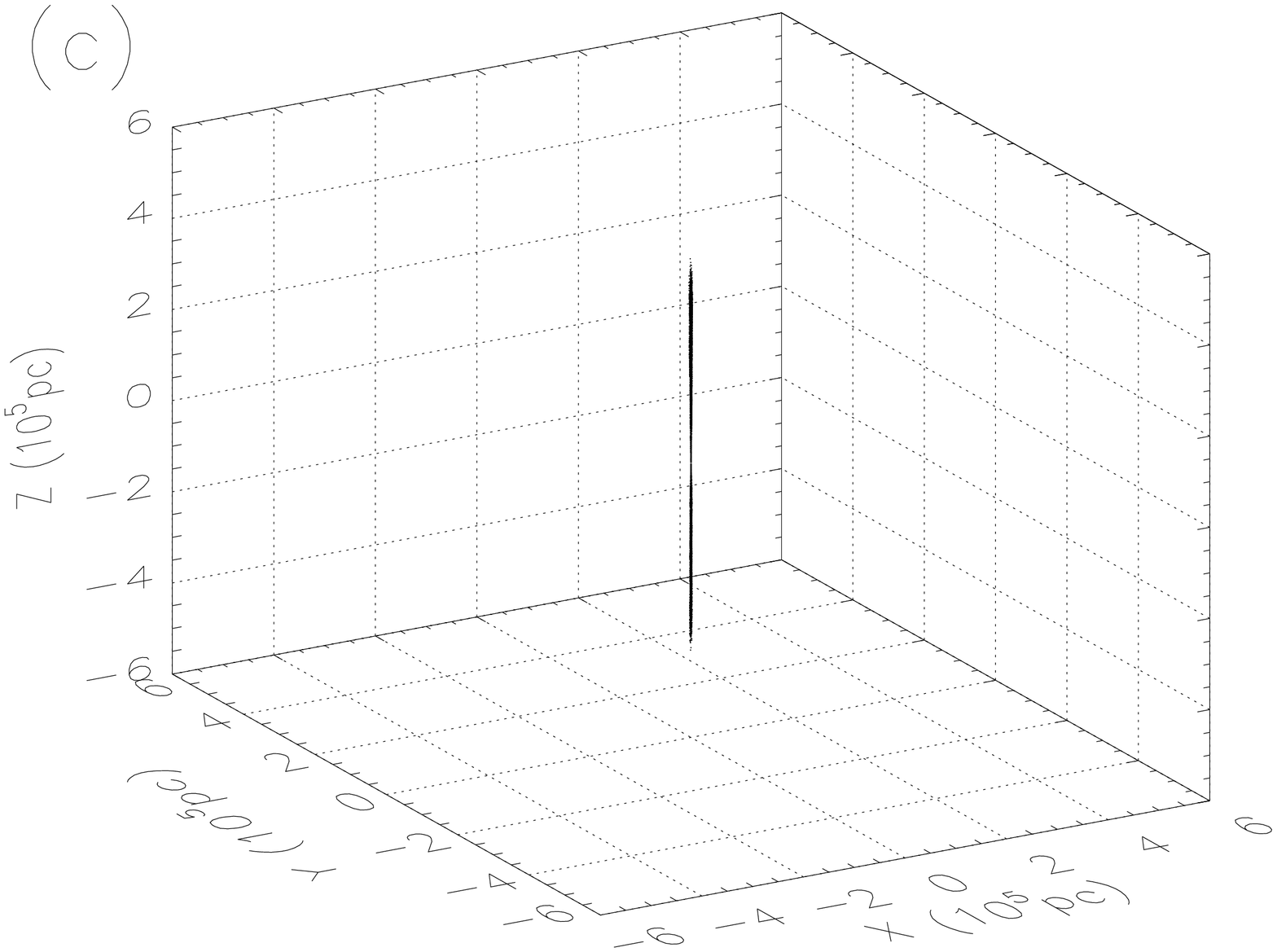}
	\caption{Positions of positrons at the end of their slowing-down 
	time in a WIM taking into account collisional transport in a 
	turbulent magnetic field ($\delta B/B_{0} = 1$ and 
	$\lambda_{\rm max}$ = 10~pc). The uniform magnetic field lies 
	along the z axis. Positrons are injected at the origin of the 
	frame with initial kinetic energies of (a) 5~keV, (b) 100~keV 
	and (c) 5MeV.\label{fig:plot3d}}
\end{figure}

We estimate the effects of the turbulent magnetic field on the distances 
traveled by positrons by adding a turbulent component to the uniform 
field. This turbulent field 
is modelled using the plane wave approximation method presented 
by~\cite{1994-Giacalone_A430}. We assume a ratio $\delta B/B_{0}\simeq 1$.
The fluctuations follow a Kolmogorov spectrum and have a maximum 
turbulent scale $\lambda_{\rm max} \simeq$ 10-100~pc in the hot 
and warm phases and $\lambda_{\rm max} \simeq$ 1-10~pc in 
the cold neutral and molecular phases. 
These scale lengths correspond to the typical sizes of the respective phases. 
Since the Larmor radii of positrons with kinetic energies in the considered 
range are extremely small compared to the turbulent scale lengths, 
positrons simply propagate along the turbulent field lines. 
Under these conditions, the actual coordinates of positrons at the end of their 
slowing-down period are obtained by carrying their coordinates along the 
uniform field (as calculated in section~\ref{sc:MC-ISM}) over 
to the curvilinear frame of the turbulent field lines. These 
coordinates are calculated by 
Monte-Carlo simulations, using a large number of randomly chosen turbulent 
configurations for a given $\lambda_{\rm max}$. Figure~\ref{fig:plot3d} 
shows examples of the positions of positrons at the end of the 
slowing-down time, calculated with this method in a WIM with 
$n_{\rm H}$ = 0.2~\rm{cm}$^{-3}$ and $f_{\rm ion}$ = 0.6. 

Adding such a turbulent magnetic field in the simulations 
reduces the FWHM of the positron field-aligned distributions by a 
factor $\simeq 0.75$ and broadens their transverse (i.e., 
perpendicular to $\vec{B_{0}}$) distributions 
due to the chaotic behavior of field lines. At low energy, 
when the distance traveled by positrons is 
shorter than the maximum scale of the turbulence, 
$\lambda_{\rm max}$ (e.g., Figure~\ref{fig:plot3d}a), positrons 
have a quasi-uniform distribution along any field line that 
can be considered, at this scale, as straight and randomly 
tilted with respect to $\vec{B_{0}}$. In this case, the 
spatial distribution of positrons is nearly
spherically symmetric. At intermediate energy 
(e.g., Figure~\ref{fig:plot3d}b), positrons have a more collimated 
distribution with a transverse dispersion that increases with 
distance from the source. At high energy, when the distance 
traveled by positrons is large compared to $\lambda_{\rm max}$
(e.g., Figure~\ref{fig:plot3d}c), the positron distribution is highly 
collimated along $\vec{B_{0}}$.

\begin{figure}[!b]
	\centering	
	\parbox{9cm}{\includegraphics*[scale=0.33]{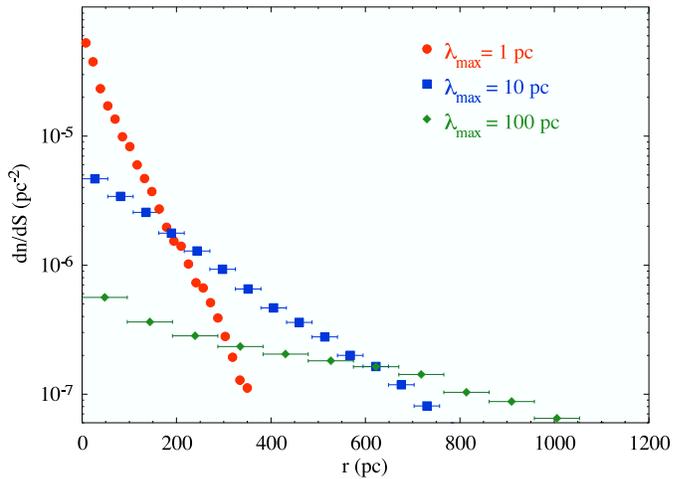}}
		\caption{Transverse distributions of positrons 
	at the end of their slowing-down period in a WIM, 
	calculated for maximum turbulent scales $\lambda_{\rm max}$ 
	= 10, 100 and 1000 pc. The initial kinetic energy of positrons is 
	E$_{0}$ = 1 MeV. The distributions are normalized to unity. 
	\label{fig:dndr-WIM}}
\end{figure}

The shape of the transverse distribution\footnote{The transverse 
distribution is $\frac{dn}{dS}$ with $dS = 2 \pi r dr$ and $r$ the 
radius in the $xy$ plane.} is strongly peaked, as illustrated in 
Figure~\ref{fig:dndr-WIM}, for positrons with $E_{0}$ = 1~MeV propagating 
in a WIM. The transverse distribution is close to a decaying exponential 
function. Its extent increases with $\lambda_{\rm max}$, as expected
since a larger scale length leads to larger transverse excursions 
of field lines away from the uniform field. 
In contrast, the extent of the field-aligned distribution does not 
change with $\lambda_{\rm max}$, because, on average, positrons propagate along 
the uniform magnetic field. 

Due to the cusped shape of the transverse distribution, 
FWHM$_{\perp}$ is not necessarily a very meaningful quantity, 
as the fraction of 
positrons inside FWHM$_{\perp}$ is low. Moreover, this fraction changes 
significantly with $\lambda_{\rm max}$ (e.g., from $\simeq$1\% to 
$\simeq$9\% for the cases presented in Figure~\ref{fig:dndr-WIM}) and 
with $E_{0}$ (e.g., from $\simeq$1\% to $\simeq$30\% for $E_{0}$ 
decreasing from 10~MeV to 1~keV). In contrast, the fraction of positrons 
inside the FWHM of the field-aligned distribution ranges from 
$\sim$ 80\% to $\sim$ 95\%.

\begin{figure}[!b]
	\centering	
	\parbox{9cm}{\includegraphics*[scale=0.33]{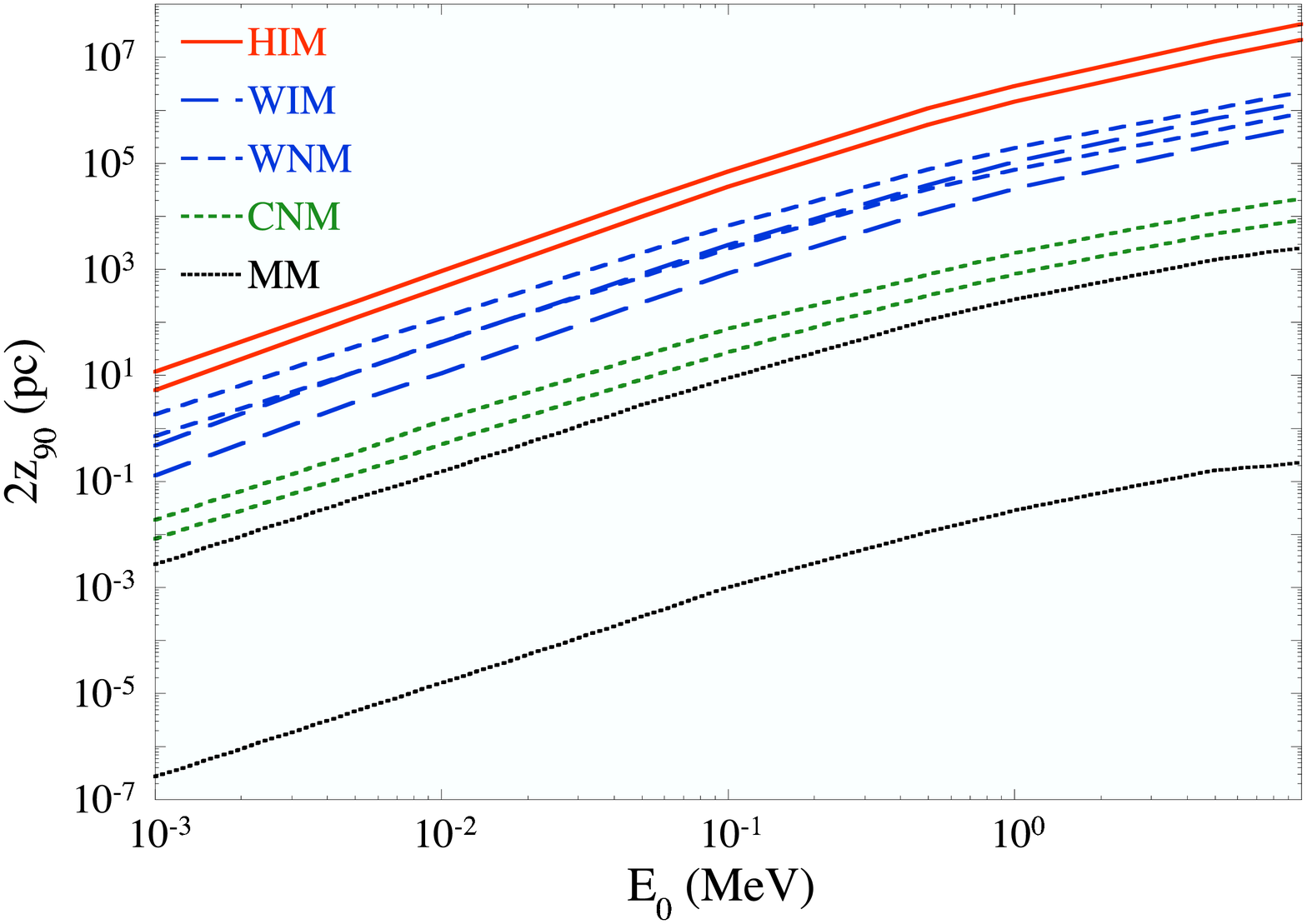}}
	\parbox{9cm}{\includegraphics*[scale=0.33]{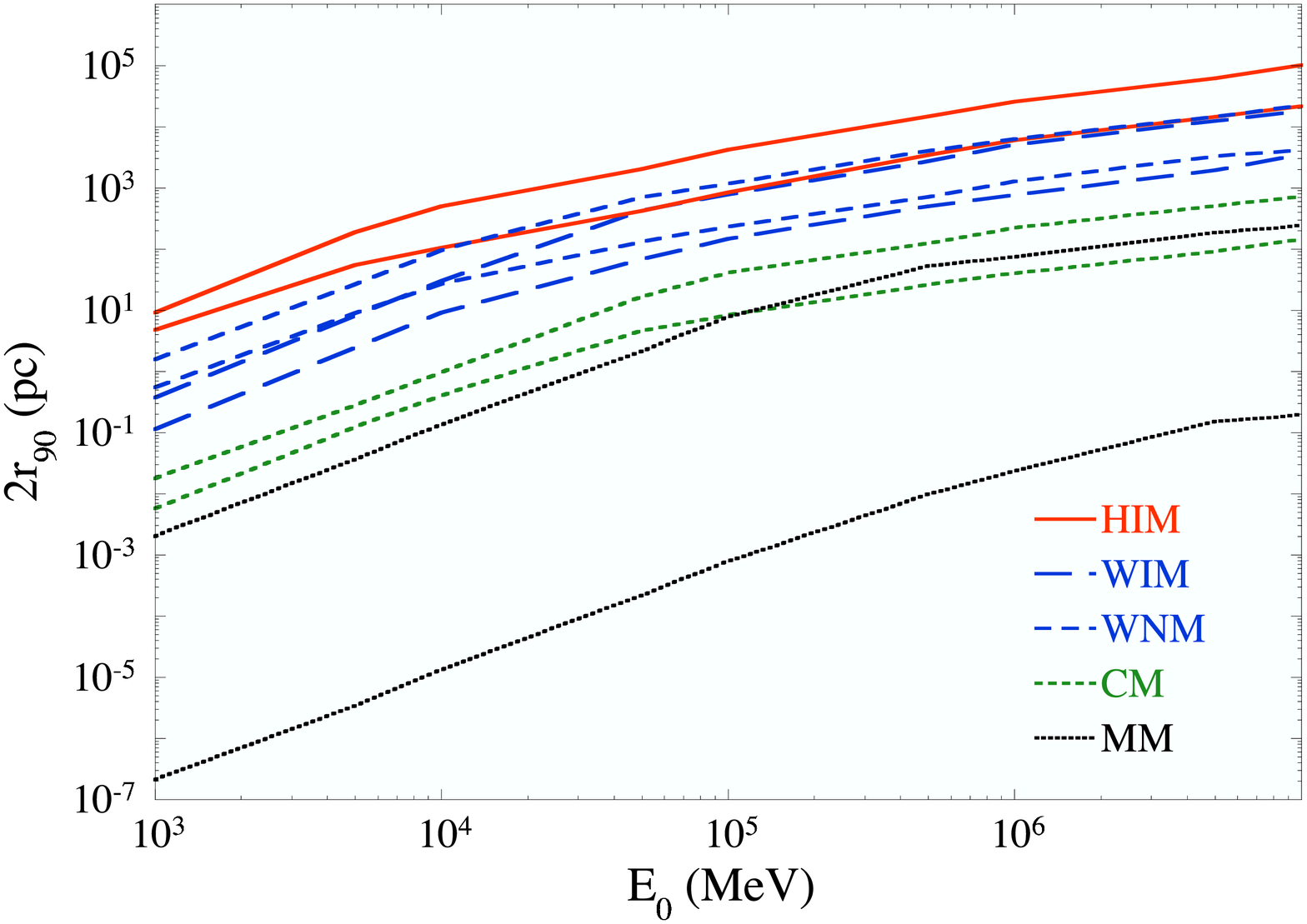}}
		\caption{Minimum and maximum extents of the spatial 
		distributions of positrons reaching 100~eV, along (top) 
		and perpendicular (bottom) to the uniform magnetic 
		field, taking into account the turbulent behavior 
		of the field lines as well as realistic values for the 
		density in each ISM phase. 
\label{fig:extenturb}}
\end{figure}

Consequently, instead of characterizing the spatial extents of the 
positron distribution by its FWHM, we will use the field-aligned 
length ($2z_{90}$) and the diameter ($2r_{90}$) that contain 90\% of 
the positrons. Figure~\ref{fig:extenturb} shows the extents 
of the spatial distribution as functions of the initial kinetic 
energy of positrons. 
The physical conditions (density, ionization fraction, temperature,\ldots) 
of the different ISM phases correspond to the extreme values 
presented in Table~\ref{T:Res}. The minimum and maximum spatial 
extents are obtained with the maximum and minimum densities and 
ionization fractions listed in Table~\ref{T:Res} and with the minimum
and maximum $\lambda_{\rm max}$, respectively. 

The ratio $r_{90}$/$z_{90}$ increases up to $\sim$0.8 with decreasing 
energy, confirming that the positron distribution is more isotropic 
at low energy. A spherically symmetric distribution would have produced 
a ratio $r_{90}$/$z_{90}$ = 1.2.  
 
The results displayed in Figure~\ref{fig:extenturb} show that MeV positrons 
travel long distances before reaching 100~eV. Positrons with $E_{0} 
\lesssim$ 30~keV generally cannot escape molecular clouds (assuming the 
smallest size 
of molecular clouds is $\sim$1 pc). On the other hand, positrons with
$E_{0} \gtrsim$ 7~keV generally escape the hot medium (assuming the typical size 
of hot regions is $\sim$200 pc). 

So far, we have discussed the distance traveled by positrons over their 
slowing-down period, which ends when their kinetic energy falls below 100~eV. 
We now consider the distance they travel before annihilating with 
electrons. In the neutral phases, low-energy positrons ($E<$100~eV) annihilate 
mostly by forming positroniums through charge exchange processes with 
atoms or molecules. The distance they travel before annihilating is negligible
compared to the distance traveled at high energy (see Table~4 of Jean et al. 
2006). The situation is different in the ionized phases. There, positrons 
quickly thermalize and then annihilate with electrons (via radiative 
recombination or direct annihilation). Since the time scale 
of annihilation ($\tau_{\rm ann}$) is inversely proportional to the electron 
density, positrons in low-density media may travel a long distance before 
annihilating. The r.m.s. distance traveled by thermalized positrons along 
a uniform magnetic field is 

\begin{equation}
    z_{\rm th} = \sqrt{2 D_{\parallel} \tau_{\rm ann}}\quad,
    \label{eq:zth}
\end{equation}
with the diffusion coefficient $D_{\parallel} = \lambda_{\rm e} 
\upsilon_{\rm e} / 3$. Here, $\upsilon_{\rm e}$ is the positron r.m.s. 
velocity, $\lambda_{\rm e}$ is the positron collisional mean 
free-path,

\begin{equation}
    \lambda_{\rm e} = (2.5 \times 10^{13} {\rm cm}) \ \frac{T^{2}_{\rm 
    e,eV}}{n_{\rm e,cm^{-3}} \, \Lambda} \quad,
    \label{eq:lcoll}
\end{equation}
and values of $\tau_{\rm ann}$ are derived from Table~3 of \cite{Guessoum_al_2005}.

The r.m.s. distance $z_{\rm th}$ characterizes the width of the Gaussian 
field-aligned distribution of thermalized positrons at the time of 
annihilation, when they are initially injected isotropically at $z$ = 0. 
Taking into account the turbulent magnetic field, the 3-D spatial distribution 
of the sites where positrons annihilate, is calculated by carrying their 
coordinates along the uniform field over to the curvilinear frame of the 
turbulent field lines. We then derive the field-aligned and transverse 
extents of this distribution ($2z_{{\rm th},90}$ and $2r_{{\rm th},90}$, 
respectively), defined as the field-aligned length and the diameter that 
contain 90\% of the annihilating positrons.
In the WIM, $z_{{\rm th},90} < $0.015~pc, i.e., $\ll z_{90}$. 
In the HIM, $z_{{\rm th},90} \sim$ 300 and $\sim$600~pc, 
for $n_{\rm e}$ = 0.01 and 0.005 cm$^{-3}$, respectively. 
These values are $> z_{90}$ for positrons with $E_{0} \lesssim$ 10~keV 
(see. Figure~\ref{fig:extenturb}). Furthermore, $z_{{\rm th},90}$ in 
the HIM is larger than the typical size of hot regions, confirming 
the estimates of \cite{Jean:2006fk} that positrons do not annihilate 
in the HIM. 

\begin{figure}[!t]
	\centering	
	\parbox{9cm}{\includegraphics*[scale=0.33]{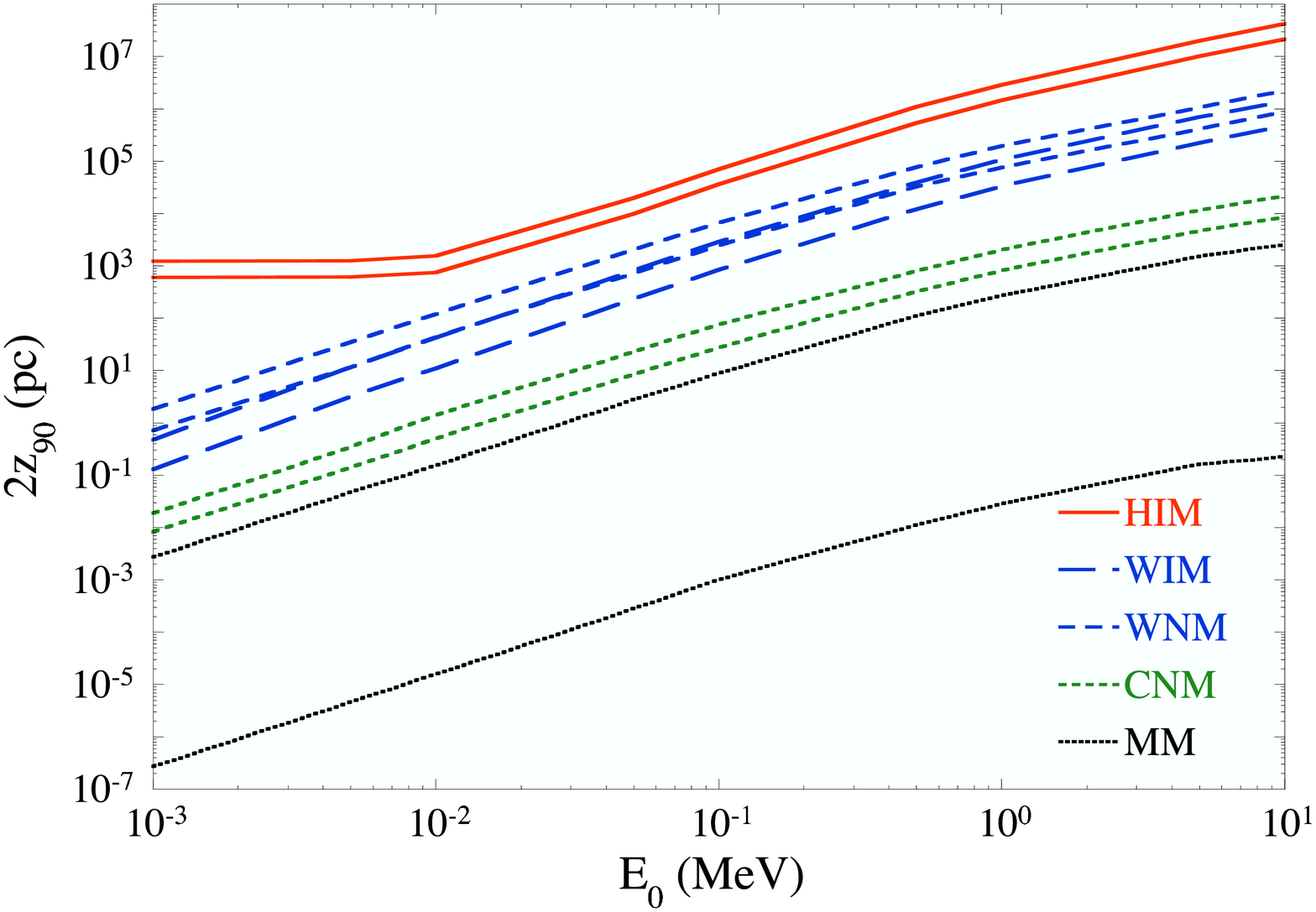}}
	\parbox{9cm}{\includegraphics*[scale=0.33]{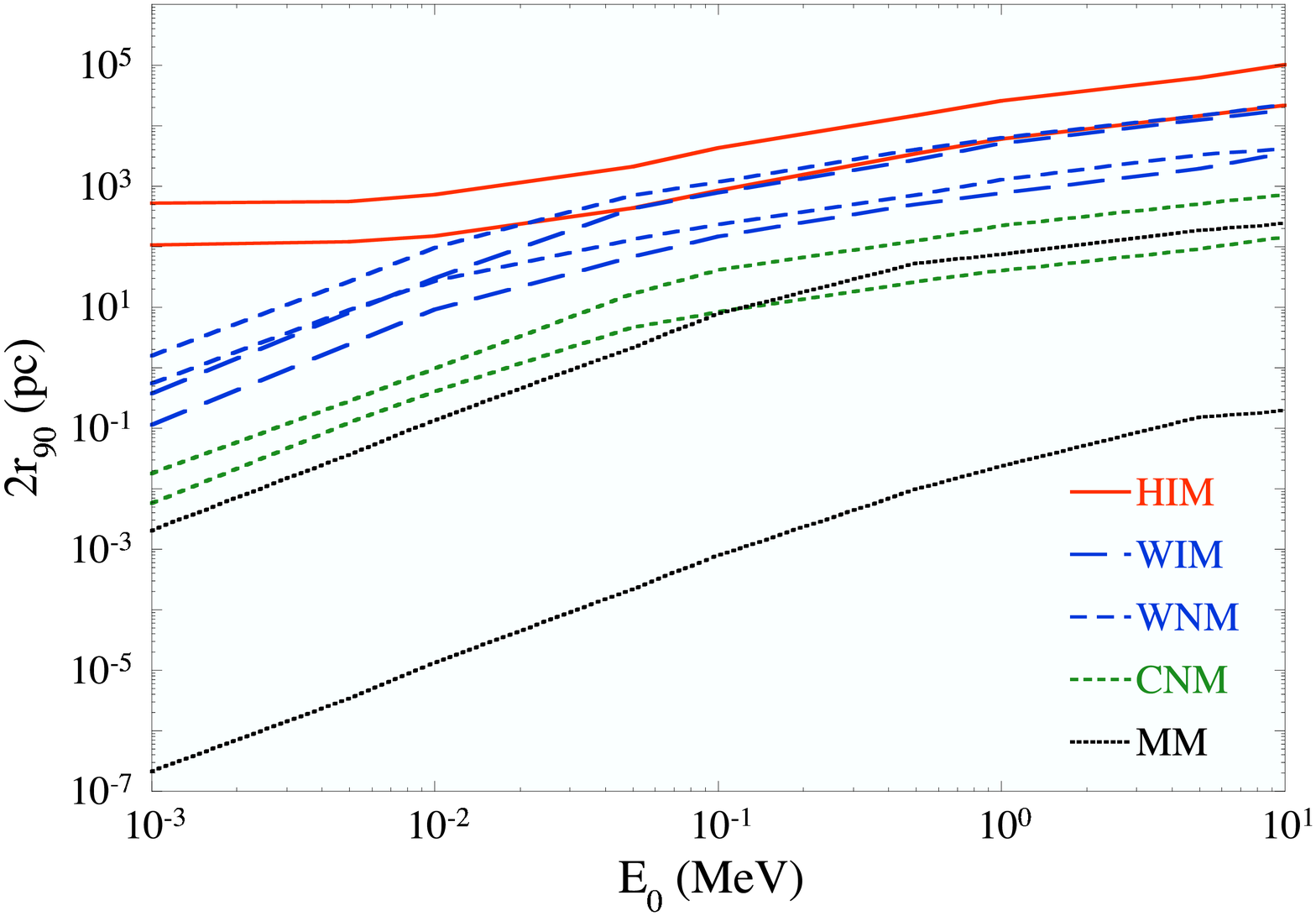}}
		\caption{Minimum and maximum extents of the spatial 
		distributions of annihilating positrons, along (top) 
		and perpendicular (bottom) to the uniform magnetic 
		field, taking into account the turbulent behavior 
		of the field lines as well as realistic values for the 
		density in each ISM phase. 
\label{fig:extenturbann}}
\end{figure}

Combining these results with the half extents ($z_{90}$ and $r_{90}$) 
presented in Figure~\ref{fig:extenturb}, we calculate the actual full 
half-extents of the spatial distribution of annihilating positrons, 
as functions of their initial kinetic energy:

\begin{equation}
    z_{{\rm ann},90} = \sqrt{z_{90}^2 + z_{{\rm th},90}^2}
    \label{eq:z90ann}
\end{equation}

and 

\begin{equation}
    r_{{\rm ann},90} = \sqrt{r_{90}^2 + r_{{\rm th},90}^2} \quad.
    \label{eq:r90ann}
\end{equation}
Figure~\ref{fig:extenturbann} presents these extents in the different phases 
of the ISM. They are similar to the results presented in Figure~\ref{fig:extenturb}, 
except in the HIM where the lifetime of thermalized positrons is so long 
that they have time to travel a non-negligible distance before they annihilate.

\section{Advection with large-scale fluid motions}
\label{S:Lsf}
 As discussed in section~\ref{sec:MHD}, low-energy positrons are at least 
loosely coupled to the thermal plasma, regardless of their
production channel ($\beta^+$ decay or cosmic-ray interactions). 
More specifically, positrons are tied to magnetic field lines 
and the latter are frozen-in into the plasma. If the plasma is subject 
to ordered or stochastic large-scale motions, positrons will be 
dragged along with the plasma and the field lines. 

The Galactic center region is the site of intense massive star formation, 
as recently confirmed by long-exposure Chandra observations
\citep{Munoetal06}; it harbours within the innermost 30 pc three bright
and young star clusters: the Arches, the Quintuplet and the central
Galactic cluster close to Sagittarius A$^*$. Most massive stars are grouped 
into clusters and/or associations of up to several hundreds.
The powerful winds from these stars together with their terminal
supernova explosions act collectively to produce hot, low-density
superbubbles \citep{MacCray87, Parizotetal04}.
These superbubbles have a profound impact on the structure of the ambient ISM.
Some of them are even powerful enough to break through the Galactic disk 
and produce chimneys, thereby venting their energy into the halo 
\citerefp{Mcclureetal06}.

Another possibility to drive strong winds in the Galaxy is again related
to the streaming instability triggered by cosmic rays escaping 
the Galactic disk into the halo. This phenomenon may be rather 
effective at distances $\sim 1$~kpc from the Galactic center, 
at the transition between the diffusion and advection zones 
\citep{Ptuskinetal97}.
In reality, the transition could well be closer to the Galactic center. 
Recent HESS observations tend to support the idea that the cosmic-ray 
energy density around the Galactic center is higher than in the solar 
system \citep{Aharonianetal06}. 

The two next subsections examine some of these aspects in more detail.

\subsection{Massive-star forming regions: small-scale advection}
\label{S:Hmsfr}
Positrons produced by radioactive decay in massive stars are
typically injected into the hot interior cavity of a superbubble.
In this hot medium, most positron transport takes place during the
slowing-down period, a phase during which positrons are subject to 
collisions with gas particles.

In addition, positrons are advected with the motions of the plasma 
filling the superbubble. Because of the presence of strong stellar winds 
and supernova shocks inside the superbubble and because of their interactions 
with each other as well as with local density inhomogeneities, 
the hot cavity is undoubtedly highly turbulent (\citeAY{Bykov01}, 
and references therein).
According to \citet{Bykov01}'s model of inhomogeneous turbulence 
in superbubbles, positrons, which follow field lines, are subject to 
the same chaotic transport, characterized by a diffusion coefficient 
$D = \upsilon_{\rm rms} \, L$. With a turbulent velocity in the range 
$\upsilon_{\rm rms} \sim (100-1000) \ {\rm km~s^{-1}}$ and a mean scale 
of turbulent eddies in the range $L \sim (1-10) \ {\rm pc}$,
corresponding to the typical distance between two stars (and two shocks),
we find $D \sim (3 \times 10^{25} - 3 \times 10^{27}) \ {\rm cm^2~s^{-1}}$.
It then follows that over their lifetime ($\tau_{\rm life} \sim 
(10^5~{\rm yr})/n_{\rm H,cm^{-3}}$, with $n_{\rm H}$ given 
in Table~\ref{T:Res}; see section~\ref{sec:COL}),
positrons diffuse over distances $\sim (80-1000) \ {\rm pc}$.
Thus chaotic motions appear to "confine" low-energy positrons in the hot medium 
rather efficiently compared to collisions and anisotropic MHD turbulence. 

Large-scale MHD compressible fluctuations also provide stochastic acceleration. 
The rate of re-acceleration is given by $\Gamma_{\rm acc} \sim 
\upsilon_{\rm rms}/9L$ 
in the limit $\upsilon_{\rm rms} \, L > \upsilon \, \lambda_{\rm mfp}$,
where $\upsilon$ is the positron velocity and $\lambda_{\rm mfp}$ 
its mean free-path controlled by pitch-angle scattering
(Bykov 2008, private communication).
Re-acceleration is efficient only if the re-acceleration rate is large
compared to the Coulomb energy loss rate 
(see Eq.~\ref{Eq:lamio}),
i.e., if $\upsilon_{\rm rms} > (80~{\rm km~s^{-1}}) \ 
(L/10~{\rm pc}) \ n_{\rm e,cm^{-3}} \ \beta^{-1}$.
Re-acceleration can contribute to delaying the annihilation of positrons.

At the scales explored by the gyromotion of positrons produced 
by radioactive decay or by cosmic-ray interactions, one should also
account for transport processes similar to those invoked for low-energy 
particles in the solar wind. 
\cite{1999-Ragot_518}, for instance, considered the role of non-resonant 
fast magnetosonic waves in the dissipation regime of turbulence,
i.e., above the steepening observed at a fraction of the thermal proton 
gyrofrequency (see \citeAY{Alexandrovaetal08}, and references therein).
This non-resonant transport process can produce efficient angular 
scattering through pitch angles $\alpha \to \pi/2$ and thus govern 
the particle mean free-path at energies $\sim 1$~MeV. 
We believe that this process adapted to the stellospheres of massive stars 
(typically a few parsecs in size) should be relevant for the problem of 
positron transport.

A full investigation of the local transport of positrons within 
the stellospheres of massive stars and of their spatial transport 
and re-acceleration in chaotic compressible turbulence is beyond 
the scope of this paper and will deserve future separate studies.

\subsection{Galactic winds: large-scale advection}
\label{S:Win}

Galactic winds can produce a systematic shift of the fluid at velocity 
$V_{\rm w}$ toward high Galactic latitudes. 
Positrons coupled to the fluid will be transported over their lifetimes 
$\tau_{\rm life}$ over typical distances $\sim V_{\rm w} \, \tau_{\rm life}$ 
(for a constant wind velocity). 
As mentioned above, large-scale winds can be driven termally 
by the collective action of massive stars (stellar winds and supernova 
explosions) in active regions of the Galaxy or by the streaming
of cosmic rays into the Galactic halo.

Several observations support the idea of starburst episodes in the history 
of our Galaxy (some of their properties were reviewed by 
\citeAY{Veilleuxetal05}). There are further pieces of evidence indicating 
an active nuclear wind in our Galaxy. 
For instance, powerful mass ejections from the Galactic center were observed 
on scales from a few arcminutes to tens of degrees at several wavelengths 
\citep{1996-Morris_A34, Yusef-Zadehetal00}. 
Some evidence for a large-scale bipolar wind from the Galactic center 
was found in infrared dust emission \citep{Bland03}. The authors 
attributed their $8.3~\rm\mu m$ observations to dust entrained in 
a Galactic wind that was powered by a central starburst several million 
years ago. 
Wind velocities in the Galactic center are difficult to estimate. 
\citet{Bland03} considered vertical velocities $\sim 150~{\rm km~s^{-1}}$ 
at Galactic heights of a few 100~pc.
A recent HST and FUSE data analysis \citep{Keeneyetal06} uncovered four 
high-velocity absorption components with typical velocities 
$\sim 250~{\rm km~s^{-1}}$ at a distance of a few kpc from the Galactic plane. 
One of these components was blueshifted and interpreted as being part of 
a Galactic fountain returning matter toward the Galactic center. 
Alternatively, these observations might correspond to wind velocities 
at high altitudes ($\gtrsim 1\rm~kpc$) that would be driven by the combined 
effect of the flux of cosmic rays escaping into the halo and the pressure 
of the thermal plasma \citep{Everettetal08}.

The wind closer to the Galactic disk or the Galactic center is certainly 
slower, and it probably speeds up with height. 
\citet{2006-Totani_P58}, instead of a starburst episode, favored a wind
with velocity $\sim 100~{\rm km~s^{-1}}$ produced by a flaring activity 
from \SgrA. Assuming convective transport (no other type of transport 
was considered in his analysis) throughout the lifetimes of positrons 
makes it possible to fill the extended 511 keV bulge.

Another constraint on wind velocities, probably more relevant to positrons 
produced in the Galactic disk, comes from the presence of cosmic-ray 
radioactive elements and the analysis of secondary-to-primary ratios. 
Incorporated into a transport model, these observables are best fitted 
with a wind of constant velocity $\lesssim 15~{\rm km~s^{-1}}$ 
\citep{Strong-et-al07, Maurinetal02}. 
Under the assumption of a constant wind velocity of $15~{\rm km~s^{-1}}$
and for typical positron lifetimes $\sim(10^5{\rm~yr})/n_{\rm H,cm^{-3}}$
(see section~\ref{sec:COL}), the distances covered by convection only
are rather modest ($\sim (1.5~{\rm pc})/n_{\rm H,cm^{-3}}$).
We do not expect this effect to be dominant in the transport of positrons 
from the Galactic disk.

\section{Conclusion}
\label{sec:App}

The purpose of this work was to carefully examine the propagation of positrons 
with kinetic energies $\lesssim$10 MeV in the different phases of the ISM. 
We identified and analyzed three main transport mechanisms: 
scattering off MHD waves, ballistic motion along magnetic field lines 
and advection with large-scale fluid motions.

In section~\ref{sec:MHD}, we presented an investigation of the 
scattering off 
of positrons by MHD waves. Two necessary conditions are required for 
this process to work:  (1) MHD waves and positron motions must satisfy a 
synchrotron resonance condition and (2) the corresponding MHD waves 
must exist. Since 
MHD waves exist only at frequencies lower than the local proton cyclotron 
frequency, the resonance condition and, therefore, the scattering off 
MHD waves are restricted to positrons with kinetic energies higher than a 
threshold which depends on the physical parameters of the local ISM 
(see Table~\ref{T:Res}). We showed that positrons in the neutral phases 
(i.e., WNM, CM and MM) do not scatter off MHD waves, because the latter 
are damped by collisional effects, causing MHD 
cascades to be truncated at scales several orders of magnitude above the 
Larmor radii of MeV positrons. In the ionized phases, Landau damping 
cuts off the Alfv\'en cascade at the dissipation scale. In principle, 
positrons could be in resonance with this part of the MHD cascade. 
However, the anisotropy of magnetic perturbations in the ISM leaves 
only a small fraction of the turbulent energy injected at large scales, 
thereby casting some doubt on this conclusion. 

We also briefly discussed the possible 
action of resonant Alfv\'en waves generated by the streaming of cosmic rays 
in the WIM. It appears that positrons can resonate with these 
waves over a rather restricted momentum range. In the case of self-generated 
streaming modes, the issue is uncertain: while the waves are probably
rapidly damped in a neutral medium, the interaction of positrons with the waves 
injected in an ionized medium may well help to confine them 
near their sources over longer timescales.

We also addressed the effect of compressional magnetosonic waves 
produced either in the large-scale cascade or injected locally, as 
in the solar wind. These waves can indeed produce 
some particle transport and re-acceleration, especially in the 
HIM. The importance of re-acceleration requires 
a kinetic treatement of the transport problem,i.e., solving a 
diffusion-convection equation.

Positrons that do not scatter off MHD waves move along magnetic field lines 
with helical trajectories which are perturbed by collisions with the neutral 
and/or charged particles of the ISM. A detailed analysis of this propagation mode, 
also called ``collisional regime'', is presented in section~\ref{sec:COL}. 
We found that the pitch angle of positrons is only slightly perturbed by 
collisions during their propagation at high energy. The spatial 
distribution of positrons at the end of their lifetimes was calculated using 
a Monte-Carlo method, assuming that positrons are isotropically emitted 
at a point source in the different ISM phases, with various characteristics of 
the turbulent magnetic field. The spatial distributions of high-energy 
positrons are nearly uniform along the mean magnetic field, but 
with an extent $\sim$1.7 times smaller than twice the maximum distance 
traveled by positrons. Turbulence induces a scattering of positrons 
perpendicular to the mean magnetic field, which cause an extent of 
the transverse distributions that is generally negligible compared 
to the extent of the field-aligned distributions, 
except for positrons at low energy (E$\lesssim$10 keV) or in high-density media 
(e.g., in the MM). The distances traveled by thermal positrons before 
they annihilate are negligible compared to the distances traveled 
at high energy, except in the HIM. There, their lifetimes are 
sufficiently long to allow them to diffuse along field lines over 
distances larger than the typical size of the HIM. 

In this collisional regime, positrons 
can travel over long distances (e.g., $\lesssim 30{\rm~kpc}/n_{\rm H,cm^{-3}}$ 
for $E_{0} = $ 1 MeV). Consequently, they go through several phases of 
the ISM before annihilating. In this case, since positrons probably 
propagate along field lines, the spatial distribution of their annihilation 
sites should depend on the spatial distribution of 
the Galactic magnetic field.

Finally, in section \ref{S:Lsf}, we reviewed the process of transport by 
large-scale fluid motions. There, positrons are coupled to magnetic 
field lines, which are frozen to the plasma. We distinguished between 
advection by star-driven motions
and by Galactic winds. In the first case, positrons experience the turbulent 
motions of the hot plasma in the winds of massive stars and in supernova 
remnants. We estimated a diffusion coefficient $\sim 3\times10^{25-27}$ cm$^2$ s$^{-1}$, 
depending on the turbulent velocity and the scale of turbulent eddies. 
In the second case, positrons are transported towards high Galactic latitudes 
by the Galactic wind, over distances $\sim V_{w} \tau_{\rm life}$. 
The Galactic wind velocity ranges from $\lesssim$15~km s$^{-1}$ 
in the Galactic disk to $\sim$200~km s$^{-1}$ a few kpc above 
the Galactic plane.

If the anisotropy of the magnetic field does not affect the scattering off 
MHD waves in the WIM and HIM, the distances traveled by positrons 
scattered off MHD waves have to be compared with those resulting 
from collisions with interstellar matter; the propagation regime 
leading to the shorter distances is the dominant one. Transport 
by winds will further increase the extent of the spatial distributions 
of positrons. Comparing the distances traveled by positrons, 
we find that scattering off MHD waves is more efficient to confine Galactic 
positrons. For instance, assuming quasi-linear diffusion, we estimate that 
90\% of the 1 MeV positrons emitted in the WIM by a point source would 
fill a sphere with radius $\sim$ 20-77~pc before being 
excluded from the resonance condition at an energy $\sim$ 
2.5-9.5~keV (see Table~\ref{T:Res}), depending on the true 
density and the maximum turbulent scale of the WIM. Low-energy positrons 
would then propagate along field lines in the collisional regime before 
annihilating. Ultimately, the spatial distribution of these positrons 
would have an extent ranging from $\sim$ 40 to 160 pc, i.e., lower than 
or similar to the typical size of the WIM. This implies that all 
or a large fraction of these positrons would annihilate in the WIM, 
where they are produced. The result is quite different in the case of 
anisotropic turbulence. If we neglect the effect of scattering off MHD 
waves in the WIM, then the spatial distribution of positrons would 
extend between $\sim$ 33~kpc and 100~kpc (see Figure~\ref{fig:extenturbann}), 
which implies that 1 MeV positrons systematically escape from the WIM.

The effect of the confinement of positrons close to their sources has 
to be taken into account in future investigations of their origin. 
The media in which positrons 
are likely to be produced are generally turbulent (supernova 
envelopes, supernova remnants, jets, winds). 
This effect would not only delay the time when 
positrons escape in the ISM, but would also lower their ``initial'' kinetic 
energy when they are released in the ISM. Therefore, such positrons 
would not propagate too far from their birth sites and the spatial 
distribution of sources would be closely related to the spatial 
distribution of the annihilation emission. Such a scenario is in favor 
of the hypothesis that LMXBs are a dominant source of positrons in 
the Galactic disk. Indeed, since these sources are expected to 
release low-energy positrons through winds and/or jets 
\citep{2006-Guessoum_457}, they may explain that the longitudinal 
asymmetry of the annihilation emission observed in the Galactic disk 
fits in with the asymmetry observed in the distribution 
of LMXBs emitting at high energy \citep{2008-Weinden-511-Assym1}. 
This requires that positrons propagate to distances 
shorter than a few kpc from their sources in the Galactic disk, in order 
to preserve the asymmetry, i.e., that their initial kinetic energies 
are lower than a few hundred keV if they propagate in the warm 
media in the collisional regime.

To conclude, we showed that the propagation of positrons in our Galaxy 
depends on their initial kinetic energy and on the physical conditions of 
the ISM in which they propagate. Therefore, in order to solve the mystery of 
the origin of Galactic positrons, a careful study and modelling of 
their sources (spatial distribution in our Galaxy, initial kinetic 
energy,\ldots) and their propagation (ambient medium,  
mean magnetic field, turbulence,\ldots) 
have to be undertaken. Here, we recall that the physical parameters 
of the different ISM phases used in this study are 
characteristic of the Galactic disk. The physical parameters are 
different in the Galactic bulge and particularly close to \SgrA{}.
A detailed calculation for each kind of source will be presented 
in a future paper. The results of such calculations will be compared to the 
spatial and spectral distributions of the annihilation emission 
measured by SPI.


\begin{acknowledgement}{}
The authors would like to thank M. Lee for his helpful comments, and 
A. Bykov for many enlighting discussions. The authors also thank the 
anonymous referee for helpful corrections and comments. This work was 
partially supported by the the ISSI (International Space Science Institute 
at Bern) programme for International Team work (ID \#110 on 
``Positron annihilation in the Galaxy'').
\end{acknowledgement}{}

\appendix
\section{MHD wave cascade damping}
\label{S:Appen}

\subsection{Collisional damping}
\label{S:Col}

\noindent \underline{Alfv\'en waves}
\begin{enumerate}
\item[a.]Ionized phases \\
With the parameter values listed in Table~\ref{T:Res}, the proton 
collisional mean free-path (Eq.~\ref{Eq:Lmfp}) is $\lambda_{\rm p} 
\simeq (0.8 - 1.6) \times 10^{18}$~cm in the hot phase and 
$\lambda_{\rm p} \simeq (1.6 - 5.8) \times 10^{12}$~cm in the warm 
ionized phase. Clearly, $L_{\rm inj} \gg \lambda_{\rm p} \gg 
k_{\pa{\rm max}}^{-1}$ in both media.

In the collisional range $L_{\rm inj} \ge k^{-1} > \lambda_{\rm p}$, 
the main damping mechanism of Alfv\'en waves is viscous damping 
\citep[e.g.,][]{Lazaretal03}, which proceeds at a rate
\beq
\label{Eq:Gamma_visc_alfven}
\Gamma_{\rm visc} = 
\frac{1}{2} \ \frac{\eta_1}{\rho_{\rm i}} \ k^2 
\left( \sin^2 \Theta + 4 \, \cos^2 \Theta \right) \ ,
\eeq
where
\beq
\label{Eq:eta1}
\eta_1 = 0.3 \ \frac{n_{\rm i} \, k_{\rm B} T_{\rm i}}
                    {\Omega_{\rm ci}^2 \, \tau_{\rm i}}
       + 0.51 \ \frac{n_{\rm e} \, k_{\rm B} T_{\rm e}}
                     {\Omega_{\rm ce}^2 \, \tau_{\rm e}}
\eeq
is the (ion + electron) viscosity coefficient 
(\citeAY{1965-Braginskii_RPP1}; note that we divided his damping 
rate, which applies to the wave energy, by a factor of 2 so as to 
obtain the amplitude damping rate). When $T_{\rm i} \simeq T_{\rm 
e}$, as will be assumed here, the electron contribution to the 
viscosity coefficient is negligible.

It then follows from Eqs.~(\ref{Eq:transfer_alfven}) and 
(\ref{Eq:Gamma_visc_alfven}) that
\beq
\label{Eq:Gamma_tau_alfven}
\Gamma_{\rm visc} \, \tau_{\rm A} \le 
0.2 \ \frac{1}{(\Omega_{\rm ci} \, \tau_{\rm i})^2} \
\frac{\upsilon_{\rm i}}{V_{\rm A}} \
(\lambda_{\rm p} \, k) \
\frac{1}{\cos \Theta} \ .
\eeq

In a fully ionized pure-hydrogen gas, the ion r.m.s. velocity, 
$\upsilon_{\rm i} = \sqrt{3 \, k_{\rm B} T_{\rm i} / m_{\rm i}}$, is 
approximately equal to the adiabatic sound speed, $C_{\rm s} = 
\sqrt{\gamma P / \rho}$.
The latter is everywhere smaller than or on the order of the Alfv\'en 
speed, $V_{\rm A}$ -- more specifically, $C_{\rm s} < V_{\rm A}$ 
everywhere, except in the hot low-$B$ phase, 
where $C_{\rm s} \simeq (2.7 - 3.8) \, V_{\rm A}$. Since, in 
addition, $\Omega_{\rm ci} \, \tau_{\rm i} \gg 1$, one has 
$\Gamma_{\rm visc} \, \tau_{\rm A} \lll 1$ throughout the collisional 
range (except in the limit $\cos \Theta \to 0$, which is of little 
interest here), so that the Alfv\'en wave cascade proceeds virtually 
undamped down to $\lambda_{\rm p}$.

\item[b.]Atomic phases \\
In the weakly ionized (warm and cold) atomic phases, Alfv\'en waves 
are damped by ion-neutral collisions \citep{Kulsrudpearce69}. Near 
the injection scale, Alfv\'en waves have $\omega = V_{\rm A,tot} \, 
k_{\pa} \ll \nu_{\rm in},\nu_{\rm ni}$, as can be verified with the 
help of the relations $\nu_{\rm in} \simeq (1.6 \times 10^{-9}~{\rm 
cm^3~s^{-1}}) \, n_{\rm n}$
and $\nu_{\rm ni} \simeq (1.6 \times 10^{-9}~{\rm cm^3~s^{-1}}) \, 
n_{\rm i}$ (see section \ref{S:Res}), together with $V_{\rm A,tot} = 
B / \sqrt{4 \pi \, m_{\rm p} \, n_{\rm H}}$, $k_{\pa} \lesssim 1 / 
L_{\rm inj}$, and thus 
$V_{\rm A,tot} \, k_{\pa} \lesssim (7.1 \times 10^{-16}~{\rm s}^{-1}) 
\, \left( B_{\rm \mu G} / \sqrt{n_{\rm H,cm^{-3}}} \right)$. Again, 
the relevant parameter values can be found in Table~\ref{T:Res}. In 
this low-frequency regime, the damping rate due to ion-neutral 
collisions is given by
\beq
\label{Eq:Gamma_in_alfven}
\Gamma_{\rm in} =
\frac{V_{\rm A,tot}^2 \, k_{\pa}^2}{2 \, \nu_{\rm ni}}
\eeq
\citep{1988-Ferriere_332}.

Introducing Eq.~(\ref{Eq:Gamma_in_alfven}) and 
Eq.~(\ref{Eq:transfer_alfven}) with $V_{\rm A} = V_{\rm A,tot}$ into 
Eq.~(\ref{Eq:cutoff_alfven}) then yields for the cutoff parallel 
wavenumber
\beq
k_{\pa{\rm cut}} 
\, = \, \frac{2 \, \nu_{\rm ni}}{V_{\rm A,tot}} 
\, = \, (1.5 \times 10^{-14}~{\rm cm}^{-1}) \ 
\frac{n_{\rm i,cm^{-3}} \, \sqrt{n_{\rm H,cm^{-3}}}}{B_{\rm \mu G}} \ 
.
\eeq
Clearly, this expression is such that $V_{\rm A,tot} \, k_{\pa} \sim 
\nu_{\rm ni} \ll \nu_{\rm in}$, which is not quite inside, but not 
too far from the validity limit of Eq.~(\ref{Eq:Gamma_in_alfven}).

A comparison with the right-hand side of Eq.~(\ref{Eq:thr_kpar_bis}) 
immediately shows that $k_{\pa{\rm cut}} \ll k_{\pa{\rm max}}$, which 
means that the Alfv\'en wave cascade is cut off by ion-neutral 
collisions way before reaching the maximum parallel wavenumber. 
Moreover, since  positrons can interact resonantly with Alfv\'en 
waves only over a restricted energy range just above $E_{\rm k,min}$, 
corresponding to a restricted wavenumber range just below $k_{\pa{\rm 
max}}$ (see section \ref{S:Res}), we may conclude that the Alfv\'en 
wave cascade will produce no waves capable of resonantly interacting 
with positrons in the (warm and cold) atomic phases of the ISM.

\item[c.]Molecular medium \\
In molecular clouds, Alfv\'en waves are again damped by ion-neutral 
collisions. Including the effect of collisions of gas particles on 
grains, \citet{Elmegreenfiebig93} accurately calculated the minimum 
scale of Alfv\'en waves in a molecular cloud of radius $R$. They 
found that the maximum value of the product $R \, k$ is on the order 
of a few for a typical cloud density $\sim 10^4~{\rm cm^{-3}}$. Since 
$R$ typically ranges between 0.01 and 10~pc, this leads to a cutoff 
wavenumber $k_{\rm cut} \sim (10^{-19} - 10^{-16})~{\rm cm^{-1}}$. 
Here, too, the smallest scale of the Alfv\'en wave cascade is 
considerably larger than the scales at which resonant interactions 
with positrons occur.\\
\end{enumerate}

\noindent \underline{Fast magnetosonic waves}
\begin{enumerate}
\item[a.]Ionized phases \\
Like Alfv\'en waves, magnetosonic waves with $L_{\rm inj} \ge k^{-1} 
> \lambda_{\rm p}$ are primarily damped by viscous 
friction\footnote{We checked that Joule damping leads to a cutoff 
wavenumber several orders of magnitude greater than $\lambda_{\rm 
p}^{-1}$. It only dominates viscous damping at propagation angles 
$\Theta \rightarrow 0$.}. However, because of their compressible 
nature, they decay away much faster than Alfv\'en waves -- by a 
factor $\sim (\Omega_{\rm ci} \, \tau_{\rm i})^2$. The correct 
expression of their viscous damping rate reads
\beq
\label{Eq:Gamma_visc_fast}
\Gamma_{\rm visc} =
\frac{1}{6} \ \frac{\eta_0}{\rho_{\rm i}} \ k^2 \ \sin^2 \Theta + 
(\Gamma_{\rm visc})_{\rm A} \ ,
\eeq
where 
\beq
\label{Eq:eta0}
\eta_0 = 0.96 \ n_{\rm i} \, k_{\rm B} T_{\rm i} \, \tau_{\rm i}
       + 0.73 \ n_{\rm e} \, k_{\rm B} T_{\rm e} \, \tau_{\rm e}
\eeq
is the (ion + electron) viscosity coefficient and $(\Gamma_{\rm 
visc})_{\rm A}$ denotes the viscous damping rate of Alfv\'en waves 
(given by Eq.~(\ref{Eq:Gamma_visc_alfven})) 
\citep{1965-Braginskii_RPP1}. The latter is completely negligible, 
except in the limit $\sin \Theta \to 0$. 

With both $(\Gamma_{\rm visc})_{\rm A}$ and the electron contribution 
to $\eta_0$ neglected, Eqs.~(\ref{Eq:transfer_fast}) and 
(\ref{Eq:Gamma_visc_fast}) combine to give 
\beq
\label{Eq:Gamma_tau_fast}
\Gamma_{\rm visc} \, \tau_{\rm F} =
0.0533 \ \frac{\upsilon_{\rm i}}{V_{\rm ms}} \
\lambda_{\rm p} \ L_{\rm inj}^{1/2} \ k^{3/2} \ \sin^2 \Theta \ .
\eeq

As mentioned below Eq.~(\ref{Eq:Gamma_tau_alfven}), $\upsilon_{\rm i} 
\simeq C_{\rm s}$, so that $\upsilon_{\rm i} / V_{\rm ms} <~1$. It 
then follows that, at the injection scale ($k = L_{\rm inj}^{-1}$), 
$\Gamma_{\rm visc} \, \tau_{\rm F} <~1$.

At the transition between the collisional and collisionless ranges 
($k = \lambda_{\rm p}^{-1}$), 
$$
\left[ \Gamma_{\rm visc} \, \tau_{\rm F} \right]_{\lambda_{\rm 
p}^{-1}} =
0.0533 \ \frac{\upsilon_{\rm i}}{V_{\rm ms}} \
\left( \frac{L_{\rm inj}}{\lambda_{\rm p}}
\right)^{1/2} \ \sin^2 \Theta \ ,
$$
which can be either smaller or larger than unity, depending on the 
considered ISM phase and on the propagation angle, $\Theta$.

In the hot phase, $\left[ \Gamma_{\rm visc} \, \tau_{\rm F} 
\right]_{\lambda_{\rm p}^{-1}} < 1$ at all propagation angles. This 
means that the entire fast magnetosonic cascade manages to reach the 
collisionless range beyond $\lambda_{\rm p}^{-1}$, with only little 
or moderate damping.

In contrast, in the warm ionized phase, $\left[ \Gamma_{\rm visc} \, 
\tau_{\rm F} \right]_{\lambda_{\rm p}^{-1}} < 1$ only for $\Theta < 
\Theta_{\rm c}$, with $\Theta_{\rm c} \simeq 2.60^\circ - 
4.15^\circ$. If all the waves preserve their propagation angles 
across the cascade, then waves with $\Theta < \Theta_{\rm c}$ reach 
the collisionless range, whereas those with $\Theta > \Theta_{\rm c}$ 
are viscous-damped before 
reaching $\lambda_{\rm p}^{-1}$, and their cutoff wavenumber, given 
by Eq.~(\ref{Eq:cutoff_fast}), is a decreasing function of $\Theta$:
\beq
\label{Eq:kcut_visc_fast}
k_{\rm cut} =
7.06 \ \left( \frac{V_{\rm ms}}{\upsilon_{\rm i}} \right)^{2/3} \
\lambda_{\rm p}^{-2/3} \ L_{\rm inj}^{-1/3} \
(\sin \Theta)^{-4/3} \ ,
\eeq
or, numerically, $k_{\rm cut} \simeq \nolinebreak \left( (0.5 - 1.0) 
\times 10^{-14}~{\rm cm^{-1}} \right) (\sin \Theta)^{-4/3}$. However, 
the actual situation may not be as clear-cut, as in reality, the 
energy transfer down the cascade is accompanied by a randomization of 
$\Theta$, due to both nonlinear interactions between modes with 
non-parallel wave vectors and wandering of magnetic field lines. 
\citet{Yanlazarian04} estimated that the variation in $\Theta$ is 
$\delta \Theta \sim (k \, L_{\rm inj})^{-1/4}$. At $k \sim 
10^{-14}~{\rm cm^{-1}}$, this gives $\delta \Theta \sim 0.02$, which 
is small, but not negligible compared to the small values of 
$\Theta_{\rm c}$. Ultimately, we may not rule out the possibility 
that even waves with $\Theta < \Theta_{\rm c}$ are cut off by viscous 
damping before reaching $\lambda_{\rm p}^{-1}$.

\item[b.]Atomic phases \\
Here, fast magnetosonic waves are damped by ion-neutral collisions, 
at a rate
\beq
\label{Eq:Gamma_in_fast}
\Gamma_{\rm in} =
\frac{V_{\rm A,tot}^2 \, k^2}{2 \, \nu_{\rm ni}} \ f(\Theta) \ ,
\eeq
with
$$
f(\Theta) = \frac{1}{2} \
\left( 1 + \frac{| V_{\rm A,tot}^2 - C_{\rm s}^2 |}
                {\sqrt{V_{\rm A,tot}^4 + C_{\rm s}^4 
                       - 2 \, V_{\rm A,tot}^2 \, C_{\rm s}^2 \, \cos 
2\Theta}}
\right)
$$
\citep{1988-Ferriere_332}.

Substitution of Eqs.~(\ref{Eq:Gamma_in_fast}) and 
(\ref{Eq:transfer_fast}) into Eq.~(\ref{Eq:cutoff_fast}) directly 
gives for the cutoff wavenumber
\beq
k_{\rm cut} = \left(\frac{2 \, \nu_{\rm ni}}{V_{\rm A,tot}^2} \ 
V_{\rm ms} \ L_{\rm inj}^{-1/2} \ \frac{1}{f(\Theta)}
\right)^{2/3} \ .
\eeq
Since $V_{\rm A,tot} > C_{\rm s}$ in both atomic phases, we may, to 
the order of the present approximation, let $V_{\rm ms} \simeq V_{\rm 
A,tot}$ and $f(\Theta) \simeq 1$ in the above equation, whereupon we 
obtain
\bea
k_{\rm cut} 
& \simeq &
\left( \frac{2 \, \nu_{\rm ni}}{V_{\rm A,tot}} \right)^{2/3} \
L_{\rm inj}^{-1/3}
\nonumber \\
& \simeq &
(0.9 \times 10^{-16}~{\rm cm}^{-1}) \
\left(
\frac{n_{\rm i,cm^{-3}} \, \sqrt{n_{\rm H,cm^{-3}}}}{B_{\rm \mu G}}
\right)^{2/3}
\ .
\eea
The fast magnetosonic wave cascade is cut off by ion-neutral 
collisions way before reaching the small scales at which positrons 
can be in resonant interaction.

\item[c.]Molecular medium \\
Our conclusion is similar to that reached for Alfv\'en waves.
\end{enumerate}

\subsection{Collisionless damping}
\label{S:Nco}
In section \ref{S:Col}, we saw that, in the ionized phases of the 
ISM, the Alfv\'en wave cascade experiences negligible collisional 
damping, which enables it to make it all the way down to the 
collisionless range $k^{-1} < \lambda_{\rm p}$. The fast magnetosonic 
wave cascade, in contrast, experiences significant collisional 
(viscous) damping. In the hot ionized phase, the collisional range 
$L_{\rm inj} \ge k^{-1} > \lambda_{\rm p}$
is sufficiently narrow that the fast magnetosonic cascade 
nevertheless reaches the collisionless range only partially 
attenuated. But in the warm ionized phase, where the collisional 
range spans roughly eight decades, the fast magnetosonic cascade 
completely (or almost completely) decays away before reaching the 
collisionless range. 

We now examine collisionless damping in the cases of interest, namely,
for the Alfv\'en cascade in the hot and warm ionized phases and for 
the
fast magnetosonic cascade in the hot ionized phase.\footnote{We will,
however, keep in mind the possibility that, in the warm ionized phase,
waves with small propagation angles might enter the collisionless 
regime.} 
In all cases, the dominant collisionless damping mechanism is linear 
Landau damping.
\bigskip

\noindent \underline{Alfv\'en waves}
\medskip

For Alfv\'en waves, an approximate expression of the linear Landau 
damping rate is \citep{Ginzburg_1961}:
\bea
\label{Eq:Landau_alfven}
\Gamma_{\rm LD} 
& = & 
\sqrt{\frac{\pi}{8}} \ \upsilon'_{\rm e} \ \frac{k^3}{k_{\rm A}^2} \ 
\frac{\cos \Theta \ \sin^2 \Theta}
     {\displaystyle \sin^2 \Theta 
      + 3 \, \frac{k^2}{k_{\rm A}^2} \, \cos^4 \Theta}
\nonumber \\
& \times & 
\left[ 
  \frac{{\upsilon'_{\rm i}}^2}{{\upsilon'_{\rm e}}^2}
+ \left( \sin^2 \Theta + 4 \, \cos^2 \Theta \right) \
  \exp \left( - \frac{V_{\rm A}^2}{2 \, {\upsilon'_{\rm i}}^2} \right)
\right] \ ,
\eea
where $\upsilon'_{\rm i} = \sqrt{k_{\rm B} T_{\rm i} / m_{\rm i}}$ 
and $\upsilon'_{\rm e} = \sqrt{k_{\rm B} T_{\rm e} / m_{\rm e}}$ are 
the ion and electron thermal speeds (which differ from the ion and 
electron r.m.s. velocities, $\upsilon_{\rm i}$ and $\upsilon_{\rm 
e}$, by a factor $\sqrt{3}$) and $k_{\rm A} = \Omega_{\rm ci} / 
V_{\rm A}$ is the inverse ion inertial length. For a pure-hydrogen 
plasma, $k_{\rm A}$ is nothing else than 
the maximum parallel wavenumber, $k_{\pa{\rm max}}$ (see 
Eq.~(\ref{Eq:thr_kpar_alfven})). 

Multiplying Eq.~(\ref{Eq:Landau_alfven}) by 
Eq.~(\ref{Eq:transfer_alfven}) immediately leads to
\bea
\label{Eq:Landau_tau_alfven}
\Gamma_{\rm LD}  \, \tau_{\rm A}
& = & 
\sqrt{\frac{\pi}{8}} \ \frac{\upsilon'_{\rm e}}{V_{\rm A}} \ 
\frac{k^2}{k_{\rm A}^2} \ 
\frac{\sin^2 \Theta}
     {\displaystyle \sin^2 \Theta 
      + 3 \, \frac{k^2}{k_{\rm A}^2} \, \cos^4 \Theta}
\nonumber \\
& \times & 
\left[ 
  \frac{{\upsilon'_{\rm i}}^2}{{\upsilon'_{\rm e}}^2}
+ \left( \sin^2 \Theta + 4 \, \cos^2 \Theta \right) \
  \exp \left( - \frac{V_{\rm A}^2}{2 \, {\upsilon'_{\rm i}}^2} \right)
\right] \ .
\eea
The cutoff of the Alfv\'en cascade occurs when $\Gamma_{\rm LD}  \, 
\tau_{\rm A} = 1$ (see Eq.~(\ref{Eq:cutoff_alfven})), provided that 
this relation admit a real solution for $k$.

In the hot low-$B$ and warm ionized phases, where $V_{\rm A} \lesssim 
\upsilon'_{\rm i}$, the first term inside the square brackets on the 
right-hand side of Eq.~(\ref{Eq:Landau_tau_alfven}) is negligible, 
and the relation $\Gamma_{\rm LD}  \, \tau_{\rm A} = 1$ admits one 
solution, given by
\beq
k_{\rm cut} \simeq
\left( \frac{8}{\pi} \right)^{1/4} \
\left( \frac{V_{\rm A}}{\upsilon'_{\rm e}} \right)^{1/2} \ 
\exp \left( \frac{V_{\rm A}^2}{4 \, {\upsilon'_{\rm i}}^2} \right) \
\left( \sin^2 \Theta + 4 \, \cos^2 \Theta \right)^{-1/2} \ 
k_{\rm A}
\eeq
(except in the limit $\sin \Theta \to 0$). Ignoring the weak 
$\Theta$-dependence of $k_{\rm cut}$ (which entails only a variation 
by a factor of 2), we find $k_{\rm cut} \simeq (0.14 - 0.18) \, 
k_{\rm A}$ in the hot low-$B$ phase and $k_{\rm cut} \simeq (0.65 - 
3.20) \, k_{\rm A}$ in the warm ionized phase. Hence, in these two 
media, the Alfv\'en cascade is cut off by Landau damping at a scale 
close to the proton inertial length.

In the hot high-$B$ phase, where $V_{\rm A}$ exceeds $\upsilon'_{\rm 
i}$ by a factor $\simeq 4.8 - 6.8$, the exponential factor on the 
right-hand side of Eq.~(\ref{Eq:Landau_tau_alfven}) becomes 
negligibly small, with the result that $\Gamma_{\rm LD}  \, \tau_{\rm 
A} < 1$ at all $k$ (except in the limit $\cos \Theta \to 0$, but in 
this limit Alfv\'en waves are viscous-damped before reaching the 
collisionless range;
see Eq.~(\ref{Eq:Gamma_tau_alfven})). If taken at face value, this 
result would lead to the erroneous conclusion that the Alfv\'en 
cascade is not cut off by Landau damping.

The truth is that our expressions for the frequency, transfer time 
and damping rate of Alfv\'en waves are all strictly valid only in the 
MHD regime, so that Eq.~(\ref{Eq:Landau_tau_alfven}) actually breaks 
down at
wavenumbers approaching the inverse proton inertial length, $k_{\rm 
A}$. The only conclusion that can be drawn from 
Eq.~(\ref{Eq:Landau_tau_alfven}) is that the cutoff of the Alfv\'en 
cascade by Landau damping occurs roughly at the inverse proton 
inertial length, $k_{\rm A}$.
\bigskip

\noindent \underline{Fast magnetosonic waves}
\medskip

For fast magnetosonic waves, linear Landau damping proceeds at a rate 
\citep{Ginzburg_1961}
\beq
\label{Eq:Landau_fast}
\Gamma_{\rm LD} =
\sqrt{\frac{\pi}{8}} \ \upsilon'_{\rm i} \ k \
\frac{\sin^2 \Theta}{\cos \Theta} \ 
\left[
  \frac{{\upsilon'_{\rm i}}}{{\upsilon'_{\rm e}}}
  + 5 \, \exp \left( - \frac{V_{\rm F}^2}
                            {2 \, {\upsilon'_{\rm i}}^2 \, \cos^2 
\Theta}
              \right)
\right] \ ,
\eeq
which, combined with Eq.~(\ref{Eq:transfer_fast}), yields
\beq
\label{Eq:Landau_tau_fast}
\Gamma_{\rm LD} \, \tau_{\rm F} =
\sqrt{\frac{\pi}{8}} \ \frac{\upsilon'_{\rm i}}{V_{\rm ms}} \
(k \, L_{\rm inj})^{1/2} \
\frac{\sin^2 \Theta}{\cos \Theta} \ 
\left[
  \frac{{\upsilon'_{\rm i}}}{{\upsilon'_{\rm e}}}
  + 5 \, \exp \left( - \frac{V_{\rm F}^2}
                            {2 \, {\upsilon'_{\rm i}}^2 \, \cos^2 
\Theta}
              \right)
\right] \ .
\eeq
The cutoff wavenumber, at which $\Gamma_{\rm LD}  \, \tau_{\rm F} = 
1$ (see Eq.~(\ref{Eq:cutoff_fast})), is given by 
\beq
\label{Eq:kcut_landau_fast}
k_{\rm cut} =
\frac{8}{\pi} \ 
\frac{V_{\rm ms}^2}{{\upsilon'_{\rm i}}^2} \
\frac{\cos^2 \Theta}{\sin^4 \Theta} \
\left[
  \frac{{\upsilon'_{\rm i}}}{{\upsilon'_{\rm e}}}
  + 5 \, \exp \left( - \frac{V_{\rm F}^2}
                            {2 \, {\upsilon'_{\rm i}}^2 \, \cos^2 
\Theta}
              \right)
\right]^{-2} \ L_{\rm inj}^{-1} \ .
\eeq
Here, the cutoff depends strongly on $\Theta$, and it diverges at
parallel propagation (just like the viscous cutoff in the warm ionized
phase; see Eq.~(\ref{Eq:kcut_visc_fast})). Once again, due to the
possible randomization of $\Theta$ (see section~\ref{S:Col}), this
mathematical divergence should not be a concern, unless the 
randomization 
process is inefficient.

The only interstellar phase that needs to be considered here is the 
hot ionized phase. In the low-$B$ case, $V_{\rm A} < \upsilon'_{\rm 
i}$ and the expression within square brackets in 
Eq.~(\ref{Eq:kcut_landau_fast}) is dominated by the exponential term 
(except in the limit $\cos \Theta \to 0$). 
Eq.~(\ref{Eq:kcut_landau_fast}) then reduces to $k_{\rm cut} \gtrsim 
(0.46 - 0.61) \, L_{\rm inj}^{-1} \, \cos^2 \Theta / \sin^4 \Theta$. 
Despite the strong $\Theta$-dependence of $k_{\rm cut}$, we may 
conclude that the fast magnetosonic cascade is severely affected by 
Landau damping
as soon as it enters the collisionless range and that it globally 
decays away shortly below $\lambda_{\rm p}$.

In the high-$B$ case, $V_{\rm A} > \upsilon'_{\rm i}$, the 
exponential term drops out from Eq.~(\ref{Eq:kcut_landau_fast}), and 
the result reads $k_{\rm cut} \simeq (1.2 - 2.3) \times 10^{5} \, 
L_{\rm inj}^{-1} \, 
\cos^2 \Theta / \sin^4 \Theta \simeq \left( (4.0 - 7.5) \times 
10^{-16}~{\rm cm}^{-1} \right) \, \cos^2 \Theta / \sin^4 \Theta$. 
Here, Landau damping is less severe, which allows the fast 
magnetosonic cascade to globally proceed over roughly three decades 
below $\lambda_{\rm p}$, before vanishing.


\nocite{1984-Dermer_280}
\bibliographystyle{aa}

\end{document}